\documentclass[onecolumn,journal]{IEEEtran}
\usepackage[table,usenames,dvipsnames]{xcolor}
\usepackage{graphicx}
\usepackage{algorithm}
\usepackage{algpseudocode}
\usepackage{array}
\usepackage[caption=false,font=normalsize,labelfont=sf,textfont=sf]{subfig}
\usepackage{textcomp}
\usepackage{stfloats}
\usepackage{url}
\usepackage{verbatim}
\usepackage{cite}
\usepackage{amsmath,amssymb,amsfonts,mathtools,bm}
\usepackage{tikz}
\usepackage{siunitx}
\usepackage{arydshln}
\usepackage{balance}
\usepackage{lipsum}
\usepackage{booktabs}
\usepackage{bbold}
\usepackage{multirow}
\usepackage{marginnote}
\usepackage{color, soul}
\usepackage{multicol}
\usepackage[makeroom]{cancel}
\usepackage{cases}
\newtheorem{definition}{Definition}[section]
\newtheorem{remark}{Remark}[section]

\usepackage{tabularx}
\usepackage{eurosym}
\usepackage{changepage}

\begin{document}

\title{Learning, fast and slow: a two-fold data-based model adaptation algorithm for processes under varying operating conditions}

\author{Laura Boca de Giuli, Alessio La Bella, and Riccardo Scattolini
\thanks{Laura Boca de Giuli, Alessio La Bella, and Riccardo Scattolini are with the Department of Electronics, Information, and Bioengineering, Politecnico di Milano, 20133 Milan, Italy (e-mails: \textsl{laura.bocadegiuli@polimi.it}, \textsl{alessio.labella@polimi.it}, and \textsl{riccardo.scattolini@polimi.it}).}}

\maketitle

\begin{abstract}
	This article proposes a novel two-fold learning framework to maintain the accuracy of data-based process models, typically affected by two types of uncertainty. \textit{Out-of-domain uncertainty} arises when the modelled process operates under conditions not represented in the training dataset, \textit{in-domain uncertainty} results from real-world plant variability. To handle out-of-domain uncertainty, a \textit{slow learning} component learns system dynamics under unexplored operating conditions; it consists of an ensemble of neural network models, featuring \textit{(i)} a combination rule that weights individual models based on the statistical proximity between their training data and the current operating condition, and \textit{(ii)} a monitoring algorithm based on statistical control charts that supervises the ensemble's performance and triggers the offline training and integration of a new model when a new operating condition is detected. To address in-domain uncertainty, a \textit{fast learning} component continuously compensates in real time for the mismatch of the slow learning model using Gaussian processes. The proposed methodology is tested on an energy process system referenced in the literature, demonstrating that the combined use of slow and fast learning components improves model accuracy compared to standard adaptation approaches.
\end{abstract}

\smallskip

\section{Introduction}
\label{sec:intro}
In recent years, data-based techniques have gained popularity for modelling complex processes due to the vast availability of data and their easier development compared to physics-based models. However, data-based models are inherently prone to plant-model mismatches, typically caused by two main sources \cite{gawlikowski2023survey}. First, \textit{out-of-domain uncertainty} \cite{yu2026transfer} arises when a model initially trained on a specific, and potentially limited, dataset is later used under system operating conditions not represented in the initial training dataset. Second, \textit{in-domain uncertainty} \cite{zhu2026uncertainty} is due to variability in real-world conditions and flaws in the chosen model architecture or training process, even when input-output data are sampled from the same operating conditions as the original training dataset. 
For instance, in complex processes such as energy systems, internal dynamics are influenced by external seasonal factors that may not be included in the initial training set \cite{wang2018novel}, causing out-of-domain uncertainty, while real-world plant variability gives rise to in-domain uncertainty. 
Note that out-of-domain and in-domain uncertainties are related to the notions of epistemic and aleatoric uncertainty \cite{der2009aleatory}, respectively. To mitigate these issues, continuous model monitoring and updating are crucial throughout the whole operational lifespan of the process \cite{behera2026sparse}. This adaptive capability, often referred to as lifelong or continual learning, is inspired by the human ability to acquire new knowledge while refining existing skills. Despite its potential, this remains a significant challenge in machine learning, as incrementally learning from non-stationary data can interfere with previously acquired knowledge, leading to the phenomenon called catastrophic forgetting \cite{parisi2019continual}. In the literature, \textit{ensemble learning} tackles out-of-domain uncertainty by combining multiple models to capture the process dynamics under different operating conditions, while \textit{online learning} addresses in-domain uncertainty by continuously updating a nominal model (or parts of it) in real time using newly acquired data.

\subsection{Related work}

\textbf{Ensemble learning:} Literature suggests that using a single model to learn significantly varying data distributions across different operating regions of process systems often leads to performance degradation \cite{zhang2012ensemble}. In contrast, ensemble learning methods \cite{mienye2022survey}, which take various forms as incremental learning \cite{van2022three} or mixture of experts \cite{kim2026explainable}, improve accuracy and generalisation by training specialised models, or experts, on specific datasets corresponding to different operating domains or tasks within a process, and combining their outputs. Different techniques can be used to combine the predictions of such models, ranging from simple arithmetic averaging of models' outputs, as discussed in \cite{wu2019machine} for neural networks (NNs), to weighted averaging, where higher weights are assigned to more accurate predictions based on past data \cite{wang2022coscl}. However, these methods can be inefficient, as they fail to account for the varying reliability of each model under different system operating conditions. To address this, methods of Bayesian model averaging combine the predictions of multiple models according to their posterior probabilities given observed data \cite{hoeting1999bayesian} and gating networks dynamically determine optimal models' weights using expectation-maximisation algorithms \cite{jacobs1991adaptive}. Nevertheless, the aforementioned approaches rely on previously encountered input-output data, making them unreliable for unexplored operating conditions. Moreover, these methods cannot be employed in predictive frameworks such as model predictive control (MPC), due to the unavailability of future output data of the system.

Beyond model combination, another challenge in ensemble learning is determining when to introduce a new model into the ensemble. To the best of our knowledge, existing literature lacks a systematic detection strategy for deciding when integrating a new model is advisable \cite{mienye2022survey,wu2019machine}.

Ultimately, while ensemble learning strategies adapt to changes in operating conditions, they fail to address real-time correction of in-domain uncertainty.

\textbf{Online learning:} Various strategies are proposed in the literature to learn online the plant-model mismatch caused by in-domain uncertainty \cite{hoi2021online}, and they generally correct either the entire model or an error-compensation component. When adapting the full model, parametric techniques such as set-membership methods are applied in MPC frameworks to estimate parameters online \cite{lorenzen2019robust}. Other approaches involve real-time parameter adaptation using the extended Kalman filter (EKF) \cite{sena2021ann}, moving horizon estimation (MHE) algorithms \cite{lowenstein2023physics}, and Bayesian neural networks (BNNs) \cite{jospin2022hands}. Moreover, 
meta-learning algorithms have been recently proposed \cite{hospedales2021meta},
where the parameters of a single model are rapidly updated when exposed to new tasks. However, these methods continuously adjust model parameters without assessing whether adaptation is necessary based on model performance. As a consequence, they may suffer from catastrophic forgetting under non-stationary operating conditions \cite{yap2021addressing}. To mitigate this issue, alternative strategies do not update model parameters but instead rely on non-parametric techniques, such as Gaussian process (GP) models, learning system dynamics based on past data \cite{scampicchio2025gaussian}. Nevertheless, their computational complexity significantly increases with the dataset size \cite{maiworm2021online}. Alternative online adaptation strategies do not learn the overall model, but rather correct the state dynamics of a physics-based nominal model with an error-compensation component updated online, for instance using GPs \cite{ hewing2019cautious}, Bayesian multi-task learning \cite{arcari2023bayesian}, or NNs \cite{dikici2025learning}. However, these strategies are only applicable when a nominal physical model is available, and not with purely black-box models, whose states typically do not have a physical interpretation.

Ultimately, whether updating the entire model or just an error-compensation component, these online strategies are unsuitable to address out-of-domain uncertainty: if adaptation is performed using each sampled data, the computational model complexity may grow to the point of intractability, as in the case of GPs. Conversely, if only a subset of data is selected, catastrophic forgetting may occur and previously encountered operating conditions may no longer be adequately represented. 

\smallskip
In summary, addressing both types of uncertainty requires a two-fold learning mechanism. To the best of our knowledge, only a few frameworks involving long-term and short-term learning mechanisms have been proposed in the literature, including dual Gaussian processes \cite{liu2025learning}, dual adaptive neural networks \cite{pham2021dualnet}, and models with different update rates \cite{aranilearning}. However, these approaches are not suited to learn multiple operating conditions as they operate within a fixed model structure whose parameters are continuously updated, possibly at different rates. Large Language Models are also emerging for this purpose \cite{zhang2025large}, but they do not explicitly distinguish between the need to adapt the model structure to a new operating condition and the need for fast online correction of in-domain uncertainty.

\subsection{Main contribution}
In view of the above discussion, the framework proposed in this work is a two-fold machine learning architecture comprising a \textit{slow learning} component that addresses out-of-domain uncertainty and a \textit{fast learning} component that addresses in-domain uncertainty. These two components involve the following main contributions:
\begin{itemize}
	\item an ensemble learning mechanism to address out-of-domain uncertainty within the slow learning component, incorporating a monitoring strategy based on multivariate control charts \cite{montgomery2009statistical}. Unlike their conventional use in statistical process control for fault detection in processes, these charts are employed to determine when a new model should be trained and integrated into the ensemble in response to a newly detected operating condition;

	\item an ensemble weighting rule in which the weights of the model outputs are not static or optimised based on past data but they are dynamically assigned according to the statistical proximity between the current operating condition and each model training data;

	\item an online error-compensation mechanism that leverages GPs not to update the state dynamics of an available physics-based model but rather to learn the plant-model mismatch of the data-based ensemble model, addressing in-domain uncertainty within the fast learning component.
\end{itemize}

Overall, this solution yields several advantages with respect to existing approaches. The slow learning component (first two contributions) tackles out-of-domain uncertainty by integrating a new model into the ensemble only when the proposed monitoring strategy detects the occurrence of a previously unexplored operating condition. This enables lifelong learning while preventing catastrophic forgetting. The statistical distance-based weighting rule allows the ensemble to continuously adapt by prioritising models trained on data that are statistically close to the current operating condition, while penalising those trained on statistically distant data. In parallel, the fast learning component (third contribution) enables real-time compensation for in-domain uncertainty that cannot be captured offline by the ensemble. Unlike existing two-fold learning methods \cite{liu2025learning,pham2021dualnet,aranilearning}, the proposed framework introduces statistically driven monitoring tools to detect unexplored operating conditions and enables event-triggered adaptation of model structure when necessary, combined with an online residual-correction mechanism. The proposed modelling architecture is tested in simulation on a referenced energy system \cite{krug2021nonlinear}, characterised by multiple operating conditions and persistent mismatch between the learned NN-based model and actual measurements. Results demonstrate that this two-fold approach outperforms conventional adaptation algorithms in terms of fitting accuracy. Preliminary works on data-based model monitoring and adaptation are proposed in \cite{de2024lifelong,de2025ensemble,de2026model}; however, they do not consider two-fold model adaptation combining offline and online learning strategies.

\subsection{Paper outline}
This article is organised as follows. Section \ref{sec:preliminaries} introduces the notation and basic definitions used throughout the paper. Section \ref{sec:procedure} presents the problem statement and the overall proposed solution, with detailed explanations of the slow and fast learning components provided in Sections \ref{sec:slow} and \ref{sec:fast}, respectively. The proposed modelling architecture is tested in simulation on a complex energy system in Section \ref{sec:results}. Final considerations are given in Section \ref{sec:conclusions}.

\section{Notation and preliminaries}
\label{sec:preliminaries}
Let $\mathbb{R}$ denote the set of real numbers and $\mathbb{R}_{\geq 0}$ the set of positive or null real numbers. Given a vector $z \in \mathbb{R}^n$, its transpose is denoted as $z^{\top}$ and its $i$-th element as $z_i$. Considering $z \in \mathbb{R}^n$ and $\alpha \in \mathbb{R}$, the equality $z = \alpha$ or the inequality $z \leq \alpha$ are intended element-wise. For a vector variable $z \in \mathbb{R}^n$ observed over $m$ time steps, i.e., $z(1), \hdots, z(m)$, where the set of time indices is $\mathcal{I}=\{1,\hdots,m\}$ with cardinality $|\mathcal{I}|$, the matrix containing these $m$ observations is denoted in bold as $\bm{z}=[z(1) \hdots z(m)] \in \mathbb{R}^{n \times m}$ and can be compactly expressed as $\bm{z}=\{z(k)\}_{\forall k \in \mathcal{I}}$. Matrices containing different vector variables observed over $m$ time steps included in $\mathcal{I}$, e.g., $\bm{z}_1, \bm{z}_2$, are indicated using calligraphic letters, i.e., $ \mathcal{D} = [ \bm{z}_1^{\top} \, \bm{z}_2^{\top} ]^{\top} $. Finally,  $I_n$ denotes the identity matrix of dimension $n \times n$, whereas $\mathbf{0}$ indicates a matrix consisting entirely of zeros.

In this paper, we make use of two statistical process control (SPC) techniques, i.e., the Mahalanobis distance and control charts, which are introduced below. Further details on SPC can be found in \cite{montgomery2009statistical}. 

\begin{definition} \label{def:T2} (\textbf{Mahalanobis distance}).
	Consider a benchmark dataset $\bm{z}=\{z(k)\}_{\forall k \in \mathcal{I}}$ containing $|\mathcal{I}|$ observations of a variable $z \in \mathbb{R}^{n_z}$, and a monitoring dataset \mbox{$\bm{\widetilde{z}}=\{z(k)\}_{\forall k \in \widetilde{\mathcal{I}}}$} with $|\widetilde{\mathcal{I}}|$ observations of $\widetilde{z} \in \mathbb{R}^{n_z}$, where $\mathcal{I}$ and $\widetilde{\mathcal{I}}$ contain the time indices of the observations in $\bm{z}$ and $\bm{\widetilde{z}}$, respectively. The statistical Mahalanobis distance of $\bm{\widetilde{z}}$ with respect to $\bm{z}$ is defined as 
	\begin{equation}
		\label{eq:T2}
		\begin{aligned}
			T^{2}(\bm{\widetilde{z}},\bm{z}) = \{(\widetilde{z}(k)-\mu_{\bm{z}})^{\top} (\Sigma_{\bm{z}})^{-1} (\widetilde{z}(k)-\mu_{\bm{z}})\}_{\forall k \in \widetilde{\mathcal{I}}}^{\top}\,,
		\end{aligned}
	\end{equation}
	where $\mu_{\bm{z}} = \frac{1}{|\mathcal{I}|}\sum\limits_{\forall k \in \mathcal{I}} \; z(k)$ and $\Sigma_{\bm{z}} = \frac{1}{|\mathcal{I}|-1}\sum\limits_{\forall k \in \mathcal{I}} \; (z(k) - \mu_{\bm{z}})(z(k) - \mu_{\bm{z}})^{\top}$.
	In detail, $T^{2}(\bm{\widetilde{z}},\bm{z}) \in \mathbb{R}^{|\widetilde{\mathcal{I}}|}_{\geq 0}$ contains a sequence of Mahalanobis distances, computed for each observation in $\widetilde{\bm{z}}$ with respect to the benchmark dataset $\bm{z}$. 
	Note that, prior to computing the Mahalanobis distance, both $\bm{\widetilde{z}}$ and $\bm{z}$ are normalised with respect to $\bm{z}$. 
\end{definition}

The second SPC tool employed in this work is \textbf{control charts}, an online process-monitoring technique used to detect assignable causes of process shifts. The control charts plot a quality characteristic against the number of samples and include an upper control limit (UCL) and a lower control limit (LCL). These limits are set so that, under normal (\textit{in-control}) conditions, nearly all sample points remain within them, whereas if a certain number of points fall outside, the process is considered \textit{out-of-control}, requiring corrective actions. These limits can be determined either theoretically using probability distributions such as the $\chi^2$- or \mbox{$F$-distribution}, or empirically from data using percentiles. Among the existing control charts, the Hotelling $T^{2}$ multivariate control chart is here considered, as it is suitable for monitoring vector variables by using as quality characteristic the Mahalanobis distance. Given $\bm{z}$ and $\bm{\widetilde{z}}$ as in Definition \ref{def:T2}, the empirical control limits of the Hotelling $T^{2}$ multivariate control chart for the monitored characteristic $T^{2}(\bm{\widetilde{z}},\bm{z})$ are
\begin{equation}
	\label{eq:controllimits}
	\begin{aligned}
		\text{LCL} = 0, \;\;\;
		\text{UCL} = p_j \, \mid \, \mathbb{P}_{\!\!e}\,(T^{2}(\bm{\widetilde{z}},\bm{z}) \leq p_j)=j/100,
	\end{aligned}
\end{equation}
where $\mathbb{P}_{\!\!e}$ denotes the empirical probability computed from the observed samples \cite{mcshane2016probability}. In this paper, we select the percentile $j=99.73$, in which case \mbox{$\mathbb{P}_{\!\!e}\,(T^{2}(\bm{\widetilde{z}},\bm{z}) \leq p_j)=99.73\%$}, as is common in SPC.

\section{Problem statement and proposed solution} 
\label{sec:procedure}
Consider a continuous-time process which, after discretisation with an appropriate sampling time $\tau$, can be represented as a discrete-time process $\mathcal{P}$ reading as
\begin{equation}
	\label{eq:system}
	\mathcal{P}: \; \left\{
	\begin{aligned}
		x_{\text{p}}(k+1) &=\; f_{\text{p}}(x_{\text{p}}(k), u(k)) \\
		y_{\text{p}}(k) \;\;&=\; g_{\text{p}}(x_{\text{p}}(k), u(k)) 
	\end{aligned},
	\right.
\end{equation}
where $f_{\text{p}}$ and $g_{\text{p}}$ are generic functions, $x_{\text{p}} \in \mathbb{R}^{n_{x_{\text{p}}}}$, $u \in \mathbb{R}^{n_u}$, and $y_{\text{p}} \in \mathbb{R}^{n_y}$ represent the state, input, and output vectors of the process, respectively, and $k$ is the adopted discrete-time index. The input $u$ includes all exogenous signals affecting $\mathcal{P}$, i.e., both manipulated control variables and external disturbances, assumed to be measurable. We also assume that the exogenous signals of $\mathcal{P}$ remain within an operating range for a certain period before transitioning to a new one, e.g., in energy systems load demand gradually transitions between seasonal operating ranges.

A dynamical model $\mathcal{M}$ of $\mathcal{P}$ is typically required for control or prediction purposes. As stated in Section \ref{sec:intro}, data-based models are inherently prone to plant-model mismatches due to out-of-domain and in-domain uncertainties. Addressing these uncertainties is crucial to develop a data-based model capable of adapting over time, similarly to the human brain's lifelong learning capability \cite{parisi2019continual}, in order to ensure reliable system simulations and effective control strategies. 

\begin{figure}
	\centering
	\includegraphics[width=0.65 \textwidth]{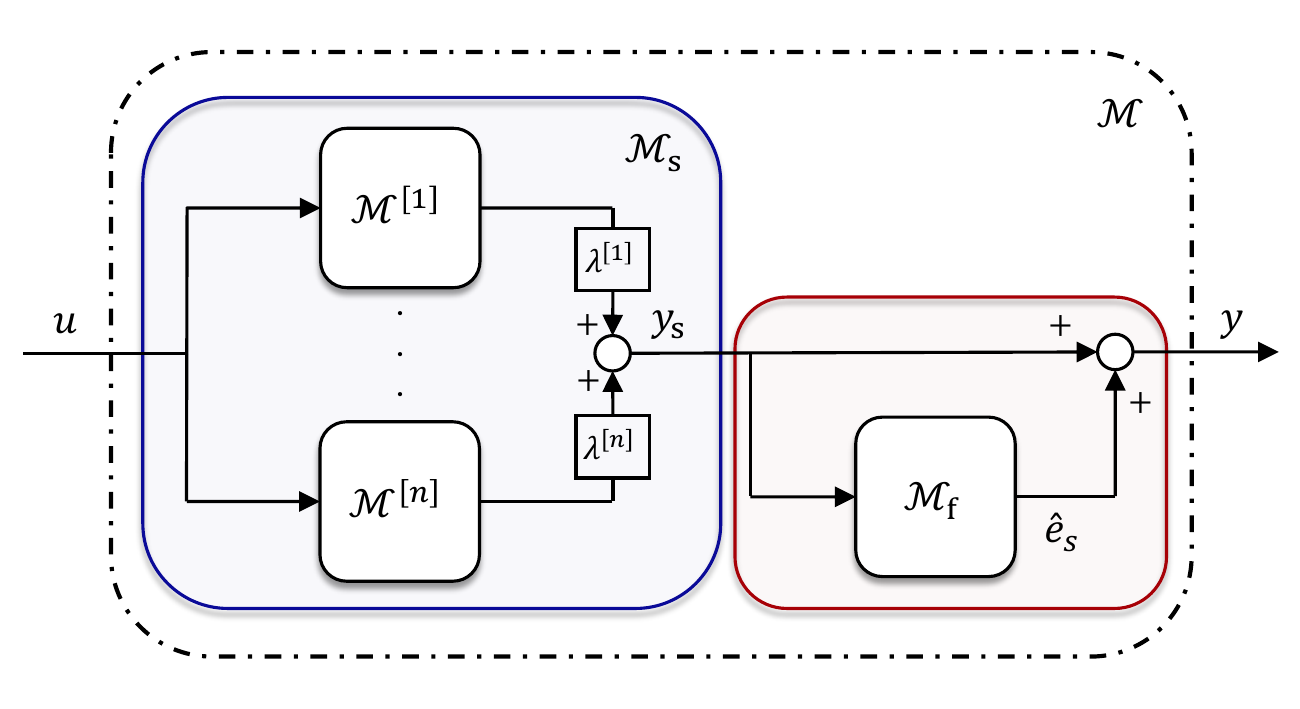}
	\caption{Proposed model architecture including slow (highlighted in blue) and fast (highlighted in red) learning.}
	\label{fig:FS}
\end{figure}

In this work, we propose a novel two-fold model architecture that enables effective adaptation to both out-of-domain and in-domain uncertainty, inspired by the dual-system theory of the Nobel Prize laureate Daniel Kahneman. According to his work ``Thinking, fast and slow'' \cite{kahneman2011thinking}, human cognition is governed by two decision-making mechanisms, referred to as Systems 1 and 2. System 2 (slow thinking) is associated with rational, slow, deliberate, analytical, cautious, and effortful reasoning. System 1 (fast thinking) is associated with irrational, fast, automatic, intuitive, and effortless reasoning, relying on heuristics and associations. Thus, the proposed architecture consists of \textit{(1)} a \textit{slow learning} component that consciously and offline tackles out-of-domain uncertainty, and \textit{(2)} a \textit{fast learning} component that automatically and online addresses in-domain uncertainty. The two components work simultaneously to improve model performance over time, each following a distinct adaptation policy. The overall model architecture, represented in Figure \ref{fig:FS}, is first presented in this section, while its individual components are detailed in Sections \ref{sec:slow} and \ref{sec:fast}.
To the best of the authors' knowledge, the parallel between the fast and slow thinking of the human brain and that of a machine learning model has not previously been explored in the literature. Existing references to the concept of ``fast and slow thinking'' in AI appear in \cite{booch2021thinking, anthony2017thinking}, where the slow process is associated with model training and the fast process with model inference, thereby offering a different interpretation of Kahneman's theory within the context of AI. 

First, slow learning consists of an ensemble learning scheme comprising a variable number of models $\mathcal{M}^{\scriptscriptstyle[1]},\hdots,\linebreak\mathcal{M}^{\scriptscriptstyle[n]}$, each trained on specific datasets $\mathcal{D}^{\scriptscriptstyle[1]},\hdots,\mathcal{D}^{\scriptscriptstyle[n]}$ corresponding to different operational domains. A new model $\mathcal{M}^{\scriptscriptstyle[n+1]}$ is trained offline and introduced into the ensemble only when deemed necessary by a model monitoring strategy. The models' outputs are combined using a convex combination depending on the current input, i.e., with weights $\lambda^{\scriptscriptstyle[i]}(u) \geq 0, \; \forall i \in \mathcal{N}=\{1,\hdots,n\}, \: \forall u \in \mathbb{R}^{n_u}$, with $\sum\limits_{i=1}^n \lambda^{\scriptscriptstyle[i]}(u)=1, \; \forall u \in \mathbb{R}^{n_u}$. This component is represented by the model $\mathcal{M}_{\text{s}}$, defined as follows:
\begin{subequations}
	\label{eq:Ms}
	\begin{numcases}{\mathcal{M}_{\text{s}}:}
		\mathcal{M}^{\scriptscriptstyle[i]}: \;
		\begin{cases}
			\label{eq:modeli}
			x^{\scriptscriptstyle[i]}(k+1) = f^{\scriptscriptstyle[i]}(x^{\scriptscriptstyle[i]}(k), u(k)) \label{eq:state_eq} \\
			y^{\scriptscriptstyle[i]}(k) = g^{\scriptscriptstyle[i]}(x^{\scriptscriptstyle[i]}(k), u(k))
		\end{cases} \\
		\begin{aligned}
			\label{eq:ys}
			\qquad\qquad y_{\text{s}}(k) = \sum\limits_{i=1}^n \lambda^{\scriptscriptstyle[i]}(u(k)) \cdot y^{\scriptscriptstyle[i]}(k)
		\end{aligned}
	\end{numcases}
\end{subequations}
where $f^{\scriptscriptstyle[i]}$ and $g^{\scriptscriptstyle[i]}$ are generic functions, \mbox{$x^{\scriptscriptstyle[i]} \in \mathbb{R}^{n_{x}^{\scriptscriptstyle[i]}}$}, $y^{\scriptscriptstyle[i]} \in \mathbb{R}^{n_{y}}$ are the state and output variables of the $i$-th model, respectively, and $y_{\text{s}} \in \mathbb{R}^{n_{y}}$ is the output of the ensemble. As will be evident in Section \ref{sec:slow}, this component takes its name from the concept of slow thinking in \cite{kahneman2011thinking}, since the offline training of new models is event-triggered, being activated only when required by the model performance monitoring strategy.

Second, fast learning consists of an online learning scheme which continuously
estimates the inherent modelling error of the slow component $\mathcal{M}_{\text{s}}$, i.e., $e_{\text{s}} = y_{\text{p}}-y_{\text{s}}$. This component is trained and updated at each time instant $k$ and is represented by the model $\mathcal{M}_{\text{f}}$, structured as follows:
\begin{equation}
	\label{eq:Mf}
	\mathcal{M}_{\text{f}}: \; \left\{
	\begin{aligned}
		x_{\text{f}}(k+1) &= f_{\text{f}}(x_{\text{f}}(k), y_{\text{s}}(k), k) \\
		\hat{e}_{\text{s}}(k) &= g_{\text{f}}(x_{\text{f}}(k))
	\end{aligned},
	\right.
\end{equation}
where $f_{\text{f}}$ and $g_{\text{f}}$ are generic functions, and \mbox{$x_{\text{f}} \in \mathbb{R}^{n_{x_{\text{f}}}}$}, $\hat{e}_{\text{s}} \in \mathbb{R}^{n_y}$ are the state and output variables, respectively, of the fast learning model. As will be evident in Section \ref{sec:fast}, this component takes its name from the concept of fast thinking in \cite{kahneman2011thinking}, since it operates in real time to refine the predictions of the slow component using recently collected information.

Equations \eqref{eq:Ms}-\eqref{eq:Mf} constitute the overall model $\mathcal{M}$, whose output is the summation of two components, i.e.,
\begin{equation}
	\label{eq:ytot}
	y(k) = y_{\text{s}}(k) + \hat{e}_{\text{s}}(k).
\end{equation}

This architecture enables to learn new operating conditions through offline training (slow learning), while ensuring real-time correction of plant-model mismatch (fast learning), allowing the overall model $\mathcal{M}$ to evolve alongside the process $\mathcal{P}$ over time.

\section{Slow learning}
\label{sec:slow}
The first decision to make in ensemble learning concerns the model structure, which is largely case study-dependent and is therefore not discussed in detail here.  Regardless of the selected model class, two additional decisions are crucial for an effective ensemble learning: how to combine the different models and when to introduce a new one.

\subsection{Combination of ensemble models}
\label{subsec:combinationrule}
The choice of the method for combining models within an ensemble is essential to achieve a reliable predictive model. In contrast to conventional strategies that employ output averaging \cite{wu2019machine} or optimisation procedures based on model outputs \cite{jacobs1991adaptive}, we propose to weight each $i$-th model output $y^{\scriptscriptstyle[i]}$ \eqref{eq:modeli} according to the statistical proximity between the input $u(k)$ and those in the training dataset $\mathcal{D}^{\scriptscriptstyle[i]}$ of $\mathcal{M}^{\scriptscriptstyle[i]}$. 

More specifically, consider an ensemble composed of $n$ models $\mathcal{M}^{\scriptscriptstyle[1]},\hdots,\mathcal{M}^{\scriptscriptstyle[n]}$, each trained on a corresponding dataset $\mathcal{D}^{\scriptscriptstyle[1]},\hdots,\mathcal{D}^{\scriptscriptstyle[n]}$. Every dataset $\mathcal{D}^{\scriptscriptstyle[i]}$ contains input-output data collected at time indices within the set $\mathcal{I}^{\scriptscriptstyle[i]}$, and is defined as $\mathcal{D}^{\scriptscriptstyle[i]} = [ \bm{u}^{\scriptscriptstyle[i] \top} \: \bm{y}_{\text{p}}^{\scriptscriptstyle[i] \top} ]^{\top}$, where \mbox{$\bm{u}^{\scriptscriptstyle[i]}=\{u(k)\}_{\forall k \in \mathcal{I}^{\scriptscriptstyle[i]}}$} and  \mbox{$\bm{y}_{\text{p}}^{\scriptscriptstyle[i]}=\{y_{\text{p}}(k)\}_{\forall k \in \mathcal{I}^{\scriptscriptstyle[i]}}$}. At each time instant $k$, the Mahalanobis distance \mbox{$T^2(u(k), \bm{u}^{\scriptscriptstyle[i]})$} between the new input $u(k)$ and the inputs in the training dataset of model $\mathcal{M}^{\scriptscriptstyle[i]}$ is computed following Definition \ref{def:T2}. The final ensemble prediction $y_{\text{s}}$ \eqref{eq:ys} is then computed through a convex combination of the outputs from the $n$ models, each weighted by
\begin{equation}
	\label{eq:lambdai}
	\begin{aligned}
		\lambda^{\scriptscriptstyle[i]}(u(k)) = \frac{w^{\scriptscriptstyle[i]}(u(k))}{\sum\limits_{i=1}^n w^{\scriptscriptstyle[i]}(u(k))}
	\end{aligned},
\end{equation}
where $w^{\scriptscriptstyle[i]}(u(k))=\frac{1}{T^2(u(k), \bm{u}^{\scriptscriptstyle[i]})+\epsilon}$ defines the statistical proximity of $u(k)$ to $\bm{u}^{\scriptscriptstyle[i]}$ and $\epsilon >0$ is an arbitrarily small constant added to avoid numerical instability when distances are close to zero. With this approach, the smaller the statistical distance \mbox{$T^2(u(k), \bm{u}^{\scriptscriptstyle[i]})$} between the new input $u(k)$ and the training inputs $\bm{u}^{\scriptscriptstyle[i]}$ of $\mathcal{D}^{\scriptscriptstyle[i]}$, the higher the weight assigned to the output of model $\mathcal{M}^{\scriptscriptstyle[i]}$, i.e., $y^{\scriptscriptstyle[i]}$. Conversely, this combination rule penalises the predictions of models whose training data are statistically distant from the current input $u(k)$. Unlike simple averaging techniques that assign equal importance to all models, this method rationally prioritises models according to the statistical proximity between the current input and their training datasets, enhancing generalisation performance even under previously unseen scenarios. Moreover, the proposed strategy involves simple calculations, i.e., \eqref{eq:T2}, making it computationally efficient for real-time evaluation. Finally, since the proposed combination rule depends only on inputs rather than outputs, it is particularly suited for predictive control frameworks such as MPC.

\subsection{Incremental learning of ensemble models}
\label{subsec:anomalydetection}
\begin{figure}
	\centering
	\includegraphics[width=0.5 \textwidth]{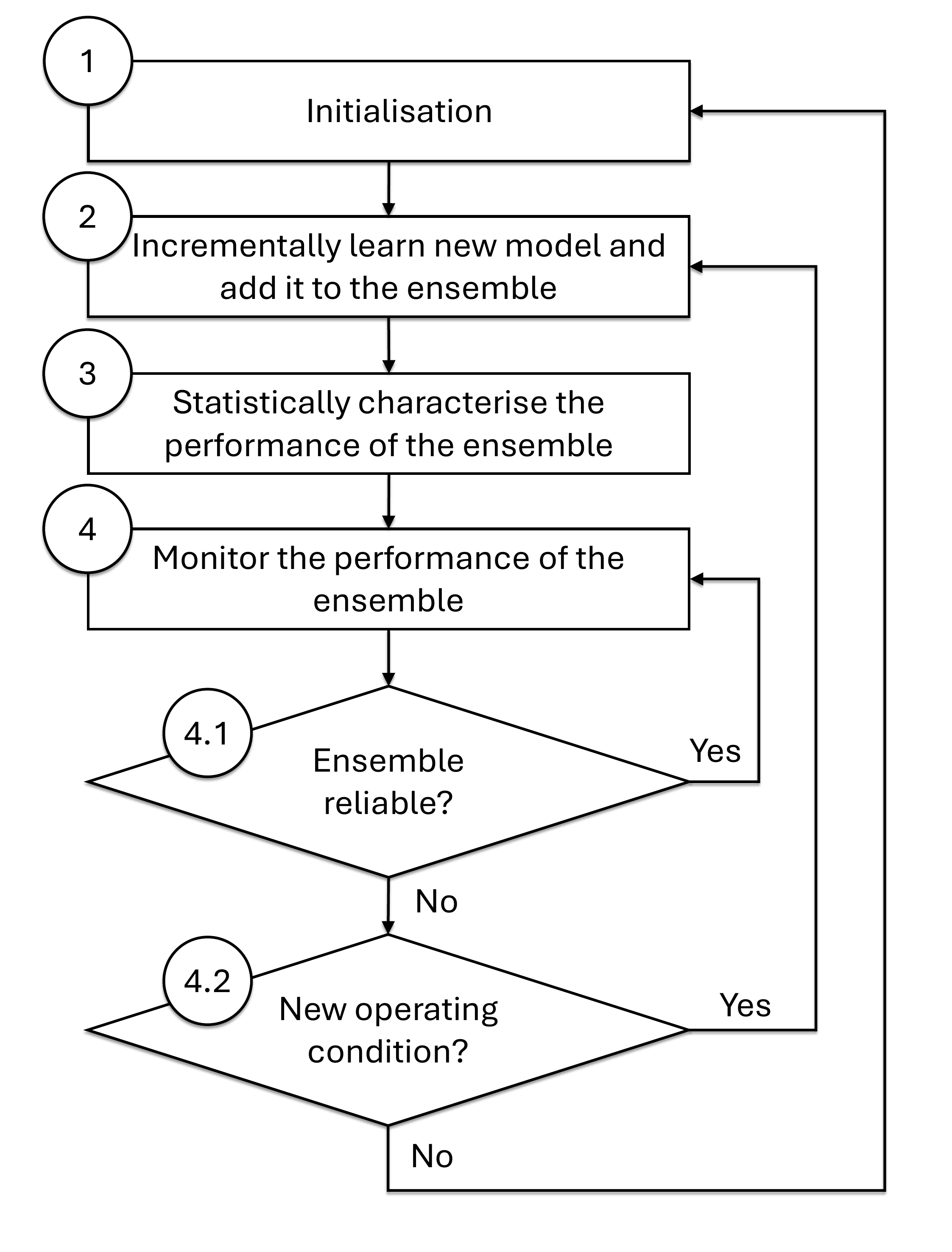}
	\caption{Flowchart of the slow learning procedure.}
	\label{fig:flowchart}
\end{figure}

A limitation of the existing ensemble learning literature is the lack
of methodologies to assess whether the overall model remains reliable
and, if not, to determine when the introduction of an additional model
becomes necessary. This motivates our additional contribution: a model performance monitoring strategy aimed at detecting anomalies in $\mathcal{M}_{\text{s}}$, identifying their causes, and implementing actions to update $\mathcal{M}_{\text{s}}$ accordingly. The proposed procedure runs iteratively during process operation, incrementally adding new models only when deemed necessary by a multivariate control chart analysis. The strategy consists of the following steps, summarised in the flowchart of Figure \ref{fig:flowchart}.

\begin{itemize}[leftmargin=12pt]
	
	\item \textbf{Step 1: Initialisation.} The number of models in the ensemble is initialised to $n=0$.
	
	\item \textbf{Step 2: Incremental model learning}. The number of models in the ensemble is increased to $n=n+1$. Then, a batch of input-output data containing $|\mathcal{I}^{\scriptscriptstyle[n]}|$ observations assumed to be under the same operating condition is collected, i.e., $\mathcal{D}^{\scriptscriptstyle[n]} = [ \bm{u}^{\scriptscriptstyle[n] \top} \: \bm{y}_{\text{p}}^{\scriptscriptstyle[n] \top} ]^{\top}$, where \mbox{$\bm{u}^{\scriptscriptstyle[n]}=\{u(k)\}_{\forall k \in \mathcal{I}^{\scriptscriptstyle[n]}}$} and \mbox{$\bm{y}_{\text{p}}^{\scriptscriptstyle[n]}=\{y_{\text{p}}(k)\}_{\forall k \in \mathcal{I}^{\scriptscriptstyle[n]}}$}. A model $\mathcal{M}^{\scriptscriptstyle[n]}$ is trained offline using this dataset.
	
	\item \textbf{Step 3: Statistical characterisation of ensemble model performance.} The ensemble model \eqref{eq:Ms}, combining the current models $\mathcal{M}^{\scriptscriptstyle[1]},\hdots,\mathcal{M}^{\scriptscriptstyle[n]}$, is characterised from a statistical perspective to establish a reference for future performance monitoring. \\	
	First, each dataset $\mathcal{D}^{\scriptscriptstyle[i]}$ collected in Step 2, $\: \forall i \in \mathcal{N}$, is divided into test and reference subsets: $\mathcal{D}^{\scriptscriptstyle[i]} = \{ \mathcal{D}^{\scriptscriptstyle[i]}_{\text{test}}, \mathcal{D}^{\scriptscriptstyle[i]}_{\text{ref}} \}$, whose time indices are contained in $\mathcal{I}^{\scriptscriptstyle[i]}_{\text{test}}$ and $\mathcal{I}^{\scriptscriptstyle[i]}_{\text{ref}}$, respectively. The modelling errors for the ensemble $\mathcal{M}_{\text{s}}$ are then collected in $\bm{e}_{\text{s},\text{test}} = \{y_{\text{p}}(k)-y_{\text{s}}(k)\}_{\forall k \in\!\! \bigcup\limits_{i \in \mathcal{N}}\!\! \mathcal{I}^{ \scriptscriptstyle[i]}_{\text{test}} }$ and in $\bm{e}_{\text{s},\text{ref}} = \{y_{\text{p}}(k)-y_{\text{s}}(k)\}_{\forall k \in \!\!\bigcup\limits_{i \in \mathcal{N}}\!\! \mathcal{I}^{ \scriptscriptstyle[i]}_{\text{ref}} }$, whereas the input data of the last learned model $\mathcal{M}^{\scriptscriptstyle[n]}$ are collected in $\bm{u}_{\text{test}}^{\scriptscriptstyle[n]}\!\!=\!\{\!u(k)\!\}_{\forall k \in \mathcal{I}^{\scriptscriptstyle[n]}_{\text{test}} }$ and in \mbox{$\bm{u}_{\text{ref}}^{\scriptscriptstyle[n]}=\{u(k)\}_{ \forall k \in \mathcal{I}^{\scriptscriptstyle[n]}_{\text{ref}} }$}.
	Using Definition \ref{def:T2}, the benchmark Mahalanobis distances $T^2(\bm{e}_{\text{s},\text{test}},\bm{e}_{\text{s},\text{ref}})$ and $T^2(\bm{u}_\text{test}^{\scriptscriptstyle[n]},\bm{u}_\text{ref}^{\scriptscriptstyle[n]})$ are computed to characterise the typical statistical distances of modelling errors and exogenous signals from their reference datasets. \\
	Finally, the benchmark Hotelling $T^{2}$ multivariate control charts are built for both $T^2(\bm{e}_{\text{s},\text{test}},\bm{e}_{\text{s},\text{ref}})$ and $T^2(\bm{u}_\text{test}^{\scriptscriptstyle[n]},\bm{u}_\text{ref}^{\scriptscriptstyle[n]})$ as described in Section \ref{sec:preliminaries}, providing the corresponding control limits UCL$_e$ and UCL$_u^{\scriptscriptstyle[n]}$, respectively (see \eqref{eq:controllimits}). 	
	
	\item \textbf{Step 4: Monitoring of ensemble model performance.} This step periodically compares operational data against the benchmark datasets $\mathcal{D}^{\scriptscriptstyle[1]},\hdots,\mathcal{D}^{\scriptscriptstyle[n]}$ to verify the reliability of the ensemble model $\mathcal{M}_{\text{s}}$ under the current operating condition. This assessment is carried out using control charts' limits, which provide a statistical threshold to determine whether the ensemble model is reliable or requires adaptation to new system dynamics.
	
	\begin{itemize}[leftmargin=16pt]
		\item \textbf{Step 4.1.} A new batch of input-output data containing $|\widetilde{\mathcal{I}}|$ observations is periodically collected, i.e., $\widetilde{\mathcal{D}} = [ \widetilde{\bm{u}}^{\top} \: \widetilde{\bm{y}}_{\text{p}}^{\top} ]^{\top}$, where \mbox{$\widetilde{\bm{u}}=\{u(k)\}_{\forall k \in \widetilde{\mathcal{I}}}$} and \mbox{$\widetilde{\bm{y}}_{\text{p}}=\{y_{\text{p}}(k)\}_{ \forall k \in \widetilde{\mathcal{I}}}$}. \\
		The modelling error of the ensemble $\mathcal{M}_{\text{s}}$ is computed as \mbox{$\widetilde{\bm{e}}_{\text{s}} = \{y_{\text{p}}(k)-y_{\text{s}}(k)\}_{\forall k \in \widetilde{\mathcal{I}} }$}. The Mahalanobis distance between the sampled errors $\widetilde{\bm{e}}_{\text{s}}$ and the reference ones $\bm{e}_{\text{s}, \text{ref}}$ is derived using Definition \ref{def:T2}, yielding \mbox{$T^2(\bm{\widetilde{e}}_{\text{s}},\bm{e}_{\text{s}, \text{ref}})$}. Leveraging the limit established in Step 3, a modelling anomaly is detected by assessing the following condition:
		\begin{equation}
			\label{eq:cond_e}
			\begin{aligned} 
				\mathbb{P}_{\!\!e}\,(T^2(\bm{\widetilde{e}}_{\text{s}},\bm{e}_{\text{s},\text{ref}}) \leq \text{UCL}_{e}) \geq j/100,
			\end{aligned}
		\end{equation}
		where $\mathbb{P}_{\!\!e}$ is the empirical probability evaluated on a batch of statistical distances and $j/100$ is typically close to 1 (see Section \ref{sec:preliminaries}).\\
		If \eqref{eq:cond_e} is verified, the ensemble modelling error remains statistically close to the scenarios explored in $\bm{e}_{\text{s},\text{ref}}$, implying that $\mathcal{M}_{\text{s}}$ is  reliable. No corrective action is required: the procedure returns to Step 4. \\
		If \eqref{eq:cond_e} is not verified, anomalous process behaviour is detected, causing out-of-control modelling errors, requiring Step 4.2.
		
		\item \textbf{Step 4.2.} To identify the source of the anomaly causing out-of-control modelling errors, we check whether the new input condition $\bm{\widetilde{u}}$ corresponds to any known operating region. \\
		To do so, \mbox{$T^2(\bm{\widetilde{u}}, \bm{u}_\text{ref}^{\scriptscriptstyle[i]})$} is computed for all $i \in \mathcal{N}$, and the following condition is checked:
		\begin{equation}
			\label{eq:cond_u}
			\begin{aligned}
				\! \exists i \in \mathcal{N} \mid  
				\mathbb{P}_{\!\!e}\,(T^2(\bm{\widetilde{u}},\bm{u}_\text{ref}^{\scriptscriptstyle[i]}) \! \leq \!  \text{UCL}_{u}^{\scriptscriptstyle[i]}) \geq j/100.
			\end{aligned}
		\end{equation} 
		If \eqref{eq:cond_u} is not verified, the current operating condition falls outside any known condition included in $\mathcal{D}^{\scriptscriptstyle[i]}_{\text{ref}}$, for all $i \in \mathcal{N}$, causing out-of-control modelling errors. This indicates that the current operating condition is statistically distant from all those used to train the existing models, implying that a new condition is detected and a new model must be identified and added to the ensemble. Step 2 is thus restored. \\
		If \eqref{eq:cond_u} is verified, it is possible to conclude that the observed inputs in $\bm{\widetilde{u}}$ are statistically close to those in a dataset $\mathcal{D}^{\scriptscriptstyle[i]}_{\text{ref}}$, with $i \in \mathcal{N}$. Thus, the current operating condition has already been explored but the current ensemble $\mathcal{M}_{\text{s}}$ no longer precisely resembles $\mathcal{P}$, as \eqref{eq:cond_e} was not verified in Step 4.1. This is a rare occurrence that may be caused by internal process changes, not captured by $\mathcal{M}_{\text{s}}$. In such cases, the ensemble model $\mathcal{M}_{\text{s}}$ is discarded, as the process has undergone internal modifications yielding anomalous modelling errors even under known operating conditions. As a result, the procedure resets at Step 1 with the training of a new ensemble.
	\end{itemize}
\end{itemize}

Overall, $\mathcal{M}_{\text{s}}$ possesses lifelong learning capabilities, as it preserves previously acquired knowledge while incorporating new information only when necessary. However, updating $y_{\text{s}}$ slowly and rationally to address out-of-domain uncertainty may not be sufficient for a complete model adaptation, as a mismatch between the process and the ensemble model $\mathcal{M}_{\text{s}}$ will inevitably occur due to in-domain uncertainty.

\section{Fast learning}
\label{sec:fast}
The fast learning component of our proposed architecture corrects online the modelling errors arising from data variability and inaccuracies in model selection or training, i.e., it compensates the ensemble model output $y_{\text{s}}$ with $\hat{e}_{\text{s}}$ so that the overall prediction $y = y_{\text{s}} + \hat{e}_{\text{s}}$, i.e., \eqref{eq:ytot}, resembles even more accurately the process output $y_{\text{p}}$. A crucial aspect of this component lies in defining 
the proper model structure. In this work, we choose a Gaussian process (GP) for $\mathcal{M}_{\text{f}}$ due to its inherent capability to adapt output predictions based on data collected online \cite{maiworm2021online}. The choice of the GP is also motivated by its non-parametric nature, which aligns with the intuitive and automatic nature of the human brain's System 1 \cite{kahneman2011thinking}, responsible for making quick decisions based on real-time stimuli. Alternative adaptive model structures, such as Bayesian neural networks \cite{jospin2022hands}, could also be employed. 

In this section, we briefly recall the fundamental concepts of GP modelling, applying them to the design of the fast learning model. For an in-depth discussion, the reader is referred to \cite{scampicchio2025gaussian}. 

Considering that the objective of $\mathcal{M}_{\text{f}}$ \eqref{eq:Mf} is to identify the dynamics of $e_{\text{s}} = y_{\text{p}} - y_{\text{s}} \, \in \mathbb{R}^{n_y}$, we leverage a GP model for each output element of $\hat{e}_{\text{s}}\, \in \mathbb{R}^{n_y}$, i.e., $\hat{e}_{\text{s},j} \, \in \mathbb{R}$, with $j \in \{1,\hdots,n_y\}$, similarly to \cite{hewing2019cautious,umlauft2017learning}. Note that, in the following, the subscript $j$, for instance in $\hat{e}_{\text{s},j}$, denotes the $j$-th element of the vector, whereas the superscript, for instance in $x_{\text{f}}^{\scriptscriptstyle[j]}$, denotes a variable associated with the $j$-th GP model. In particular, each $j$-th GP model takes as input the corresponding output of the ensemble model $\mathcal{M}_{\text{s}}$, i.e., $y_{\text{s},j}$, and employs a non-linear autoregressive model with exogenous input (NARX) \cite{maiworm2021online}. The state associated with the $j$-th GP model is composed as
\begin{equation}
	\label{eq:GPstate}
	\begin{aligned}
		x_{\text{f}}^{\scriptscriptstyle[j]}(k) = [ e_{\text{s},j}(k) \hdots e_{\text{s},j}(k-n_{r,e}+1) \, y_{\text{s},j}(k-1) \hdots y_{\text{s},j}(k-n_{r,y}) ]^{\top},
	\end{aligned}
\end{equation}
where $n_{r,e}$ and $n_{r,y}$ are the regression horizons employed. To simplify the presentation, we introduce an auxiliary variable including the state and the input of the $j$-th GP model, i.e., $\nu^{\scriptscriptstyle[j]}(k) = [ x_{\text{f}}^{\scriptscriptstyle[j]}(k)^{\top} \: y_{\text{s},j}(k)]^{\top}$.
At this point, we define the dataset \mbox{$\mathcal{D}_k^{\scriptscriptstyle[j]} = [\bm{\nu}^{\scriptscriptstyle[j]^{\top}} \: \bm{e}_{\text{s}}^{\scriptscriptstyle[j]^{\top}}]^{\top}$} employed by the $j$-th GP model, where \mbox{$\bm{\nu}^{\scriptscriptstyle[j]} \!=\! \{\nu^{\scriptscriptstyle[j]}(h)\}_{\forall h=0,\hdots,k-1}$} and \mbox{$\bm{e}_{\text{s}}^{\scriptscriptstyle[j]} = \{e_{\text{s},j}(h)\}_{\forall h=1,\hdots,k}$}. Each $j$-th GP model predicts the $j$-th output using the current state, the input, and the collected dataset, i.e., $\hat{e}_{\text{s},j}(k+1)  = f_{\text{GP}}^{\scriptscriptstyle[j]}(\nu^{\scriptscriptstyle[j]}(k)|\mathcal{D}^{\scriptscriptstyle[j]}_k)$. A GP model is fully specified by its mean and covariance functions: we here assume no prior knowledge is available and thus set the mean function to zero \cite{umlauft2017learning}.
Ultimately, according to GP modelling, the prediction of the $j$-th output element at time instant $k$ is given by the posterior mean function, defined as:
\begin{equation}
	\label{eq:GP}
	\begin{aligned}
		\hat{e}_{\text{s},j}(k+1) = f_{\text{GP}}^{\scriptscriptstyle[j]}(\nu^{\scriptscriptstyle[j]}(k)|\mathcal{D}^{\scriptscriptstyle[j]}_k) = \Sigma_1^{\scriptscriptstyle[j]}(\nu^{\scriptscriptstyle[j]}(k), \bm{\nu}^{\scriptscriptstyle[j]}) \cdot (\Sigma_2^{\scriptscriptstyle[j]}(\bm{\nu}^{\scriptscriptstyle[j]}, \bm{\nu}^{\scriptscriptstyle[j]}))^{-1} \cdot \bm{e}_{\text{s}}^{\scriptscriptstyle[j]},
	\end{aligned}
\end{equation}
where $\Sigma_1^{\scriptscriptstyle[j]} \in \mathbb{R}^{1 \times k}$ is
\begin{equation}
	\label{eq:sigma1}
	\begin{aligned}
		\Sigma_1^{\scriptscriptstyle[j]}(\nu^{\scriptscriptstyle[j]}(k), \bm{\nu}^{\scriptscriptstyle[j]}) = [ \sigma^{\scriptscriptstyle[j]}(\nu^{\scriptscriptstyle[j]}(k),\nu^{\scriptscriptstyle[j]}(0)) \hdots  \sigma^{\scriptscriptstyle[j]}(\nu^{\scriptscriptstyle[j]}(k),\nu^{\scriptscriptstyle[j]}(k-1))],
	\end{aligned}
\end{equation}
whereas $\Sigma_2^{\scriptscriptstyle[j]} \in \mathbb{R}^{k \times k}$ is composed by elements
\begin{equation}
	\label{eq:sigma2}
	\begin{aligned}
		(\Sigma_2^{\scriptscriptstyle[j]})_{h_1,h_2} \! = \! \sigma^{\scriptscriptstyle[j]}(\nu^{\scriptscriptstyle[j]}(h_1),\nu^{\scriptscriptstyle[j]}(h_2)), \forall h_1,h_2 \in \{0,\hdots,k-1\}
	\end{aligned},
\end{equation}
and $\sigma^{\scriptscriptstyle[j]}(\cdot,\cdot)$ is the covariance function associated with each output element $j \in \{1,\hdots,n_y\}$. In this work, we consider the squared exponential kernel as covariance function \cite{hewing2019cautious}, which, for two generic observations $\nu^{\scriptscriptstyle[j]}(h_1)$ and $\nu^{\scriptscriptstyle[j]}(h_2)$, reads as:
\begin{equation}
	\label{eq:kernelfunction}
	\begin{aligned}
		\sigma^{\scriptscriptstyle[j]}(\nu^{\scriptscriptstyle[j]}(h_1),\nu^{\scriptscriptstyle[j]}(h_2)) =  \alpha^{\scriptscriptstyle[j]^2} e^{-1/2((\nu^{\scriptscriptstyle[j]}(h_1)-\nu^{\scriptscriptstyle[j]}(h_2))^{\top} L^{\scriptscriptstyle[j]}(\nu^{\scriptscriptstyle[j]}(h_1)-\nu^{\scriptscriptstyle[j]}(h_2)))},
	\end{aligned}
\end{equation}
where $\alpha^{\scriptscriptstyle[j]}$ and $L^{\scriptscriptstyle[j]}$ are determined for each $j \in \{1,\hdots,n_y\}$ by maximising the log marginal likelihood using the dataset $\mathcal{D}_k^{\scriptscriptstyle[j]}$, following Bayesian inference principle \cite{maiworm2021online}. 
Ultimately, considering $e_{\text{s}} =[e_{s,1} \hdots e_{s,n_y}]^{\top}$, we can define the overall state of $\mathcal{M}_{\text{f}}$ as 
\begin{equation}
	\label{eq:GPstatetot}
	\begin{aligned}
		x_{\text{f}}(k) = [ e_{\text{s}}(k)^{\top} \hdots e_{\text{s}}(k-n_{r,e}+1)^{\top}   y_{\text{s}}(k-1)^{\top} \hdots  y_{\text{s}}(k-n_{r,y})^{\top} ]^{\top},
	\end{aligned}
\end{equation}
where $x_{\text{f}}\in\mathbb{R}^{n_{x_{\text{f}}}}$, with ${n_{x_{\text{f}}}}=n_{y}(n_{r,y}+n_{r,e})$. Moreover, we introduce the auxiliary variable $\nu(k) = [ x_{\text{f}}(k)^{\top} \: y_{\text{s}}(k)^{\top}]^{\top}$, and the dataset $\mathcal{D}_k = [\bm{\nu}^{\top} \: \bm{e}_{\text{s}}^{\top}]^{\top}$, with \mbox{$\bm{\nu} = \{\nu(h)\}_{\forall h=0,\hdots,k-1}$} and \mbox{$\bm{e}_{\text{s}} = \{e_{\text{s}}(h)\}_{\forall h=1,\hdots,k}$}. In this way, Equations \eqref{eq:GP}-\eqref{eq:kernelfunction} can be compactly written considering each element $\hat{e}_{\text{s},j}$, with $j \in \{1,\hdots,n_y\}$, as $\hat{e}_{\text{s}}(k+1) = f_{\text{GP}}(\nu(k)|\mathcal{D}_k)$.
Overall, the model $\mathcal{M}_{\text{f}}$, initially introduced in \eqref{eq:Mf}, can now be explicitly formulated as
\begin{equation}
	\label{eq:MfGP}
	\mathcal{M}_{\text{f}}: \left\{
	\begin{aligned}
		x_{\text{f}}(k+1) = [ & f_{\text{GP}}(\nu(k)|\mathcal{D}_k)^{\!\top}  e_{\text{s}}(k)^{\top} \! \hdots \! e_{\text{s}}(k-n_{r,e}+2)^{\!\top} y_{\text{s}}(k)^{\top} \hdots y_{\text{s}}(k-n_{r,y}+1)^{\top}]^{\top} \\
		\hat{e}_{\text{s}}(k) = \: \, \, & C_{\text{f}} \, x_{\text{f}}(k)\end{aligned}
	\right.
\end{equation} 
with $C_{\text{f}} = [I_{n_y} \,\mathbf{0}]\in \mathbb{R}^{{n_y}\times n_{x_{\text{f}}}}$. Once the structure of the fast learning model is defined, we introduce the algorithm that leverages it. The proposed procedure, outlined in Algorithm \ref{algo:2}, runs continuously during online process operation. After initialisation, at each time step $k$, the algorithm checks whether sufficient data are available for reliable GP training ($k_{\text{min}}$) and whether the maximum number of data samples has been reached ($k_{\text{max}}$), updating the training dataset accordingly. The GP model is then trained in real time to learn the modelling error of the data-based ensemble output.
\begin{algorithm}
	\caption{Fast learning algorithm}
	\label{algo:2}
	\begin{algorithmic}[1]
		\medskip
		\State Initialise $k=0$, the modelling error prediction $\hat{e}_{\text{s}}(k) = 0$, the minimum and maximum number of data samples to train the GP, i.e., $k_{\text{min}}$ and $k_{\text{max}}$
		
		\smallskip
		\State At each $k$: 
		\Statex \:\:\: Compensate the output of $\mathcal{M}_{\text{s}}$ as $y(k) = y_{\text{s}}(k) + \hat{e}_{\text{s}}(k)$
		\Statex \:\:\: Collect $y_{\text{p}}(k)$, $y_{\text{s}}(k)$ and compute $e_{\text{s}}(k) = y_{\text{p}}(k) - y_{\text{s}}(k)$
		\Statex \:\:\: \textbf{if} $k < k_{\text{min}}$ (not enough data for GP training):
		\Statex \:\:\: \qquad Set $\hat{e}_{\text{s}}(k+1)=\hat{e}_{\text{s}}(k)$ and go to Step 2
		\smallskip
		\Statex \:\:\: \textbf{else} (sufficient data for GP training):
		\smallskip
		\Statex \:\:\: \qquad \textbf{if} $k < k_{\text{max}}$ (less than maximum data samples):
		\Statex \:\:\: \qquad \qquad \vspace{-0.4cm}
		\begin{adjustwidth}{1.4cm}{0cm}
			Introduce $\mathcal{D}_k^{\scriptscriptstyle[j]}=[ \bm{\nu}^{\scriptscriptstyle[j]^{\top}} \: \bm{e}_{\text{s}}^{\scriptscriptstyle[j]^{\top}}]^{\top} \; \forall j \in \{1,\hdots,n_y\}$, with $\bm{\nu}^{\scriptscriptstyle[j]} = \{\nu^{\scriptscriptstyle[j]}(h)\}_{\forall h=0,\hdots,k-1}$, $\bm{e}^{\scriptscriptstyle[j]}_{\text{s}} = \{e_{\text{s},j}(h)\}_{\forall h=1,\hdots,k}$
		\end{adjustwidth}
		\smallskip
		\Statex \:\:\: \qquad \textbf{else} (maximum data samples reached):
		\Statex \:\:\: \qquad \qquad \vspace{-0.4cm} 
		\begin{adjustwidth}{1.4cm}{0cm}
			Introduce \mbox{$\mathcal{D}_k^{\scriptscriptstyle[j]}=[ \bm{\nu}^{\scriptscriptstyle[j]^{\top}} \bm{e}_{\text{s}}^{\scriptscriptstyle[j]^{\top}}]^{\top} \forall j \! \in \! \{1,\hdots,n_y\}$}, with \mbox{$\bm{\nu}^{\scriptscriptstyle[j]} = \{\nu^{\scriptscriptstyle[j]}(h)\}_{\forall h=k-k_{\text{max}},\hdots,k-1}$} and \mbox{$\bm{e}^{\scriptscriptstyle[j]}_{\text{s}} = \{e_{\text{s},j}(h)\}_{\forall h=k-k_{\text{max}}+1,\hdots,k}$}
		\end{adjustwidth}
		
		\medskip
		\State Train the GP model
		\Statex \:\:\: \vspace{-0.4cm}
		\begin{adjustwidth}{0.3cm}{0cm} For each $j \in \{1,\hdots,n_y\}$ train the kernel function \eqref{eq:kernelfunction} with $\mathcal{D}_k^{\scriptscriptstyle[j]}$ by maximising the log marginal likelihood
		\end{adjustwidth}
		
		\medskip
		\State Compute the GP prediction
		\Statex \:\:\: \vspace{-0.4cm}
		\begin{adjustwidth}{0.3cm}{0cm} For each $j \in \{1,\hdots,n_y\}$ compute $\hat{e}_{\text{s},j}(k+1)$ using $\mathcal{D}_k^{\scriptscriptstyle[j]}$ in Equation \eqref{eq:GP}
		\end{adjustwidth}
		\Statex \:\:\: Go to Step 2
	\end{algorithmic}
\end{algorithm}

\begin{remark}
	\label{remark:GPdataset}
	In Algorithm \ref{algo:2}, we limit the maximum number of data samples used to train the GP to $k_{\text{max}}$ by adding new data while discarding the oldest ones, as the computational complexity of GP regression increases with the number of data \cite{maiworm2021online}: see \cite{scampicchio2025gaussian} for other data-limiting strategies.
\end{remark}

\begin{remark}
	In this work, the error-compensation model is structured in a multiplicative (or serial) configuration, i.e., it is fed by the output of the model to be corrected (see \eqref{eq:Mf}). Alternative structures like the additive (or parallel) configuration can also be employed \cite{1102555}.
\end{remark}

Overall, relying solely on online error correction to adjust model outputs may be insufficient. Indeed, the fast learning component, like most real-time adaptation strategies, operates on a limited dataset to ensure computational efficiency during online execution, e.g., collected over a few hours in energy and process systems. In contrast, the slow learning component requires a sufficiently large batch of data to carry out a reliable offline identification procedure, e.g., collected over several hours or days. 
In the following section, we demonstrate that addressing both out-of-domain and in-domain uncertainties with the combined use of slow and fast learning leads to a substantial improvement in model accuracy compared to conventional adaptation methods.

\section{Numerical results} 
\label{sec:results}
\begin{figure}
	\centering
	\includegraphics[width=0.6 \textwidth]{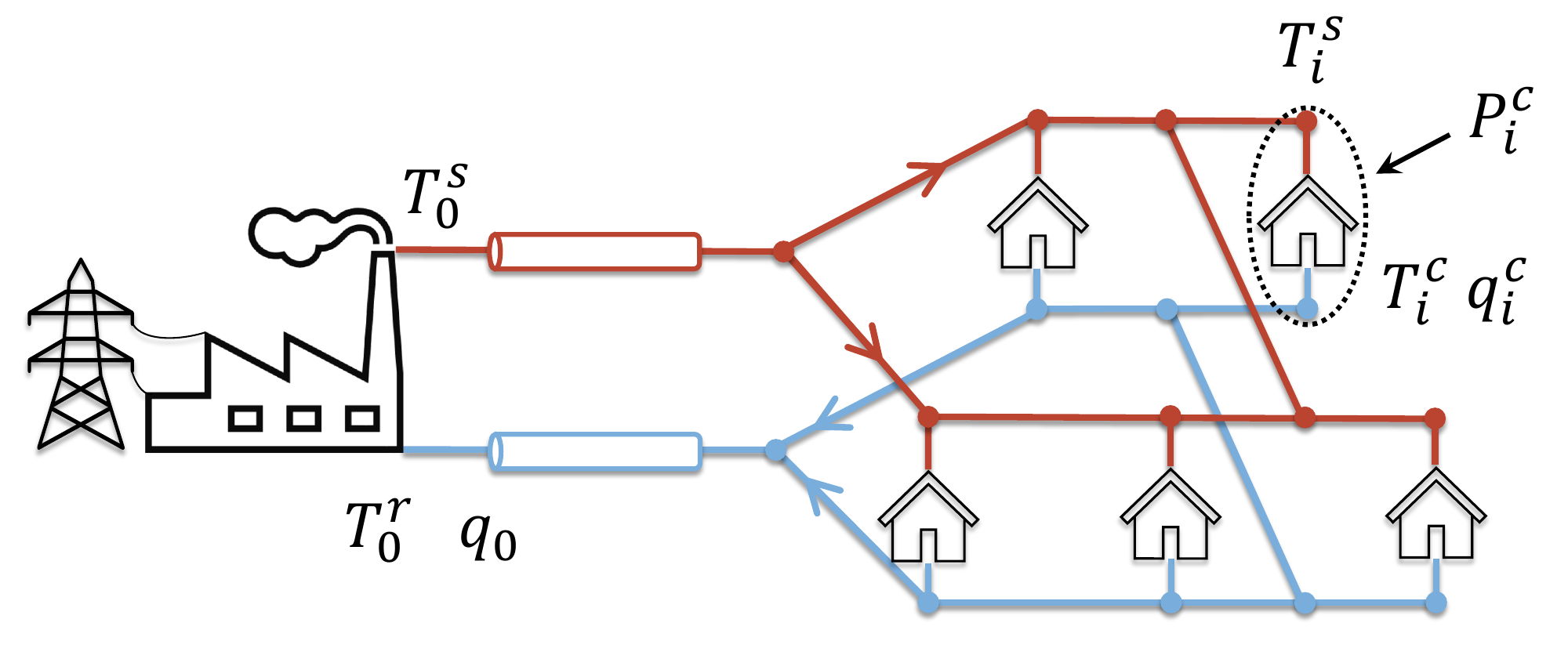}
	\caption{Schematic representation of the AROMA DHS and its variables.}
	\label{fig:aroma}
\end{figure}

The proposed modelling framework is tested on a district heating system (DHS) from the literature, i.e., the AROMA DHS presented in \cite{krug2021nonlinear} and depicted in Figure \ref{fig:aroma}. Its physical model is leveraged to develop a high-fidelity dynamic simulator based on a dedicated Modelica library \cite{alvarado2024development}, which serves as a digital twin to generate input-output data for training and evaluating the data-based modelling framework. 
The following system variables are considered as inputs and outputs for the data-based model, as shown in Figure \ref{fig:aroma} and described in \cite{de2024physics}. The control variable is the supply temperature at the heating station, denoted as \mbox{$T_0^s$}, while the external disturbances are the five thermal load demands, i.e., $P_i^c$, with $i=1,\hdots,5$. The overall input vector is thus defined as $u = [T_0^s \; P_1^c \, \hdots \, P_5^c ]^{\top} $. The output variables include the return temperature $T_0^{r}$ and the water flow $q_0$ at the heating station, as well as the supply temperature $T_i^{s}$, output temperature $T_i^{c}$, and water flow $q_i^{c}$ for each $i$-th thermal load. The corresponding output vector is thus given by \mbox{$y_p = [ T_0^{r} \; q_0 \; T_1^s \, \hdots \, T_5^s \; T_1^c \, \hdots \, T_5^c \; q_1^c \, \hdots \, q_5^c]^{\top}$}. Overall, the plant has $n_u = 6$ inputs, $n_y = 17$ outputs, and it is governed by non-linear dynamics, making it a complex process to model with physical laws and to identify from data.  

To assess the effectiveness of the proposed two-fold modelling architecture, the case study is presented following a realistic scenario where new operating conditions emerge incrementally and the model is updated accordingly. In particular, the performance of the model's slow learning component is first presented, followed by a discussion of the online compensation provided by the fast learning algorithm, and finally a comparison of the proposed method with existing adaptation approaches. All computations are performed on a laptop equipped with an Intel Core i7-10750H processor, using Python 3.12 to identify offline the data-based models and MATLAB R2024b for the monitoring and integration of the fast and slow learning frameworks.

\subsection{Slow learning results}
\vspace{2pt}

\textit{\textbf{Identification of $\bm{\mathcal{M}^{\scriptscriptstyle[1]}}$ and control charts characterisation}}: The identification process for the AROMA DHS model starts from scratch, i.e., the number of models is set to $n=0$, in line with Step 1 of the slow learning procedure (see Section \ref{sec:slow}-B). 

\begin{figure}
	\centering	\captionsetup[subfloat]{labelfont=scriptsize,textfont=scriptsize}
	\subfloat[]{ \includegraphics[width=0.4\textwidth]{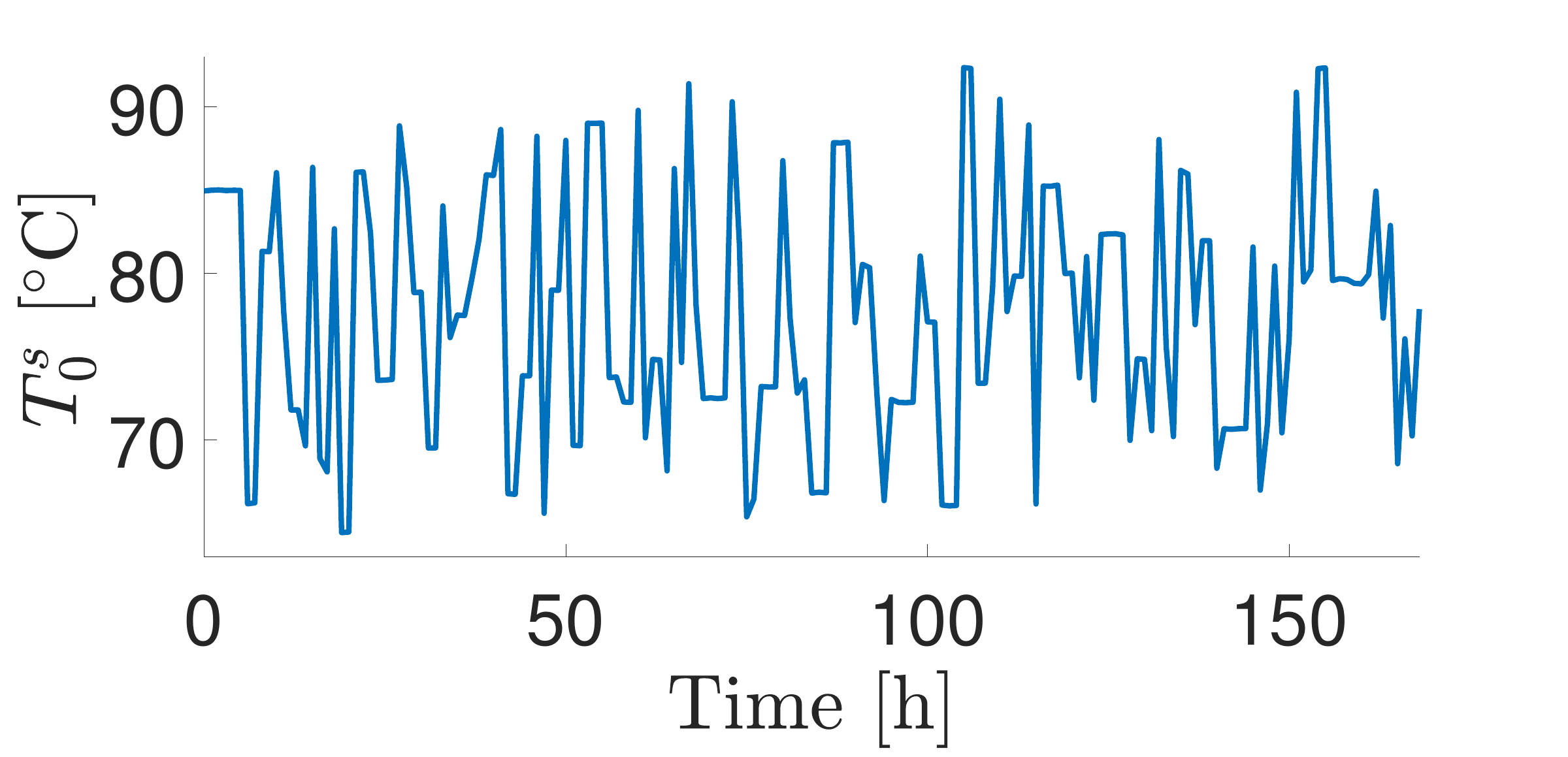} }
	\subfloat[]{\includegraphics[width=0.4\textwidth]{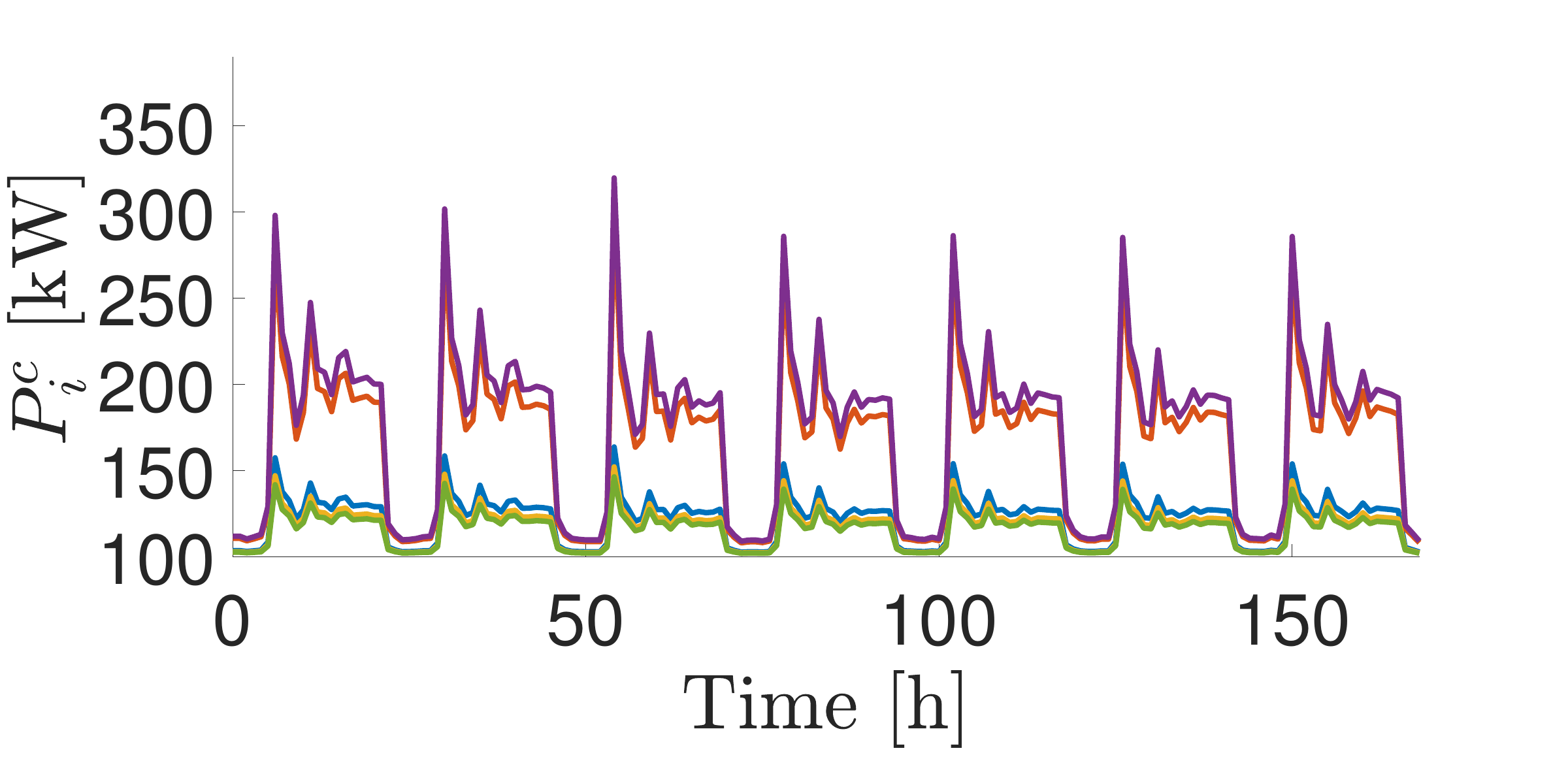} } \\
	\subfloat[]{ \includegraphics[width=0.4\textwidth]{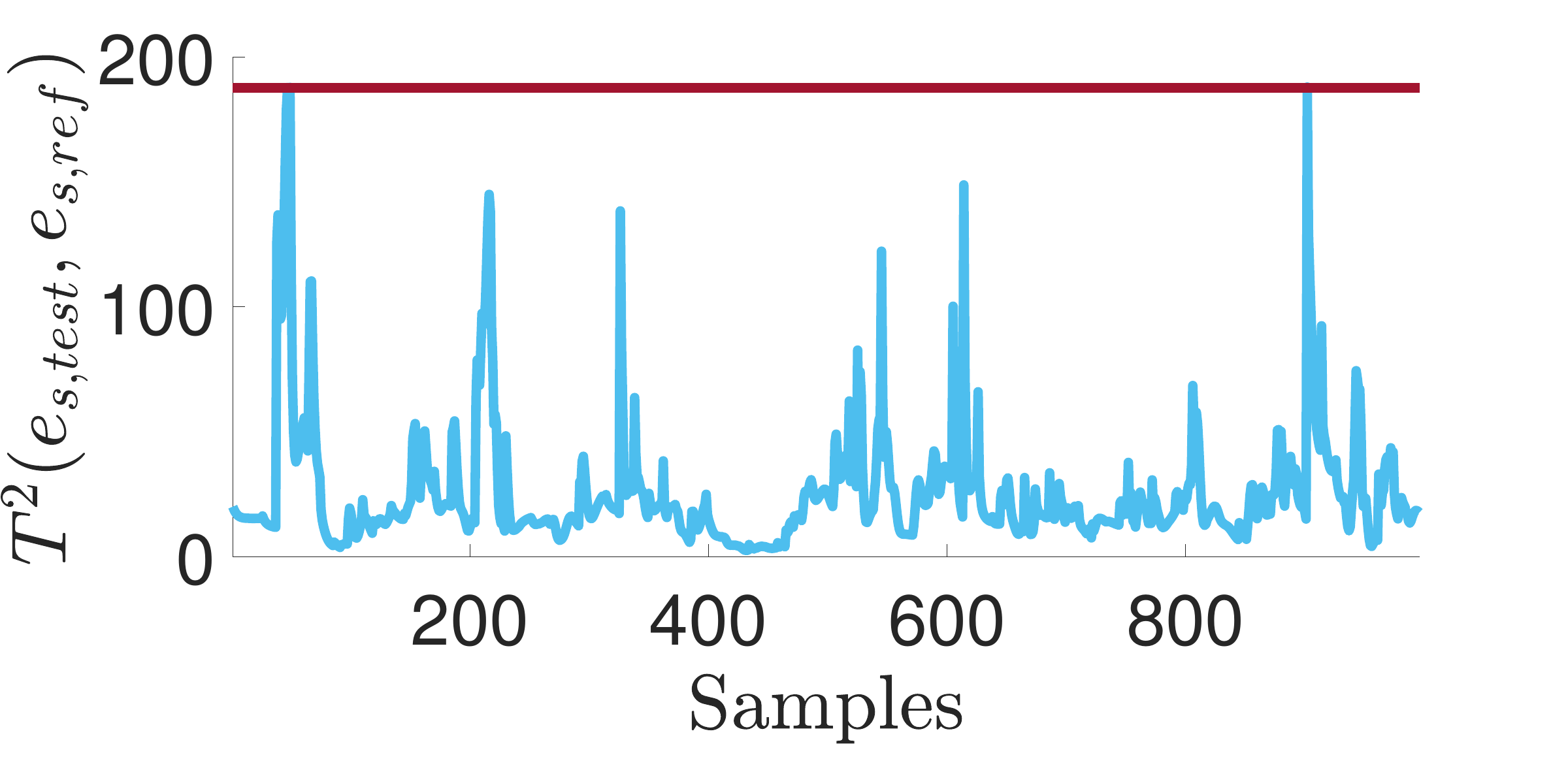} }
	\subfloat[]{\includegraphics[width=0.4\textwidth]{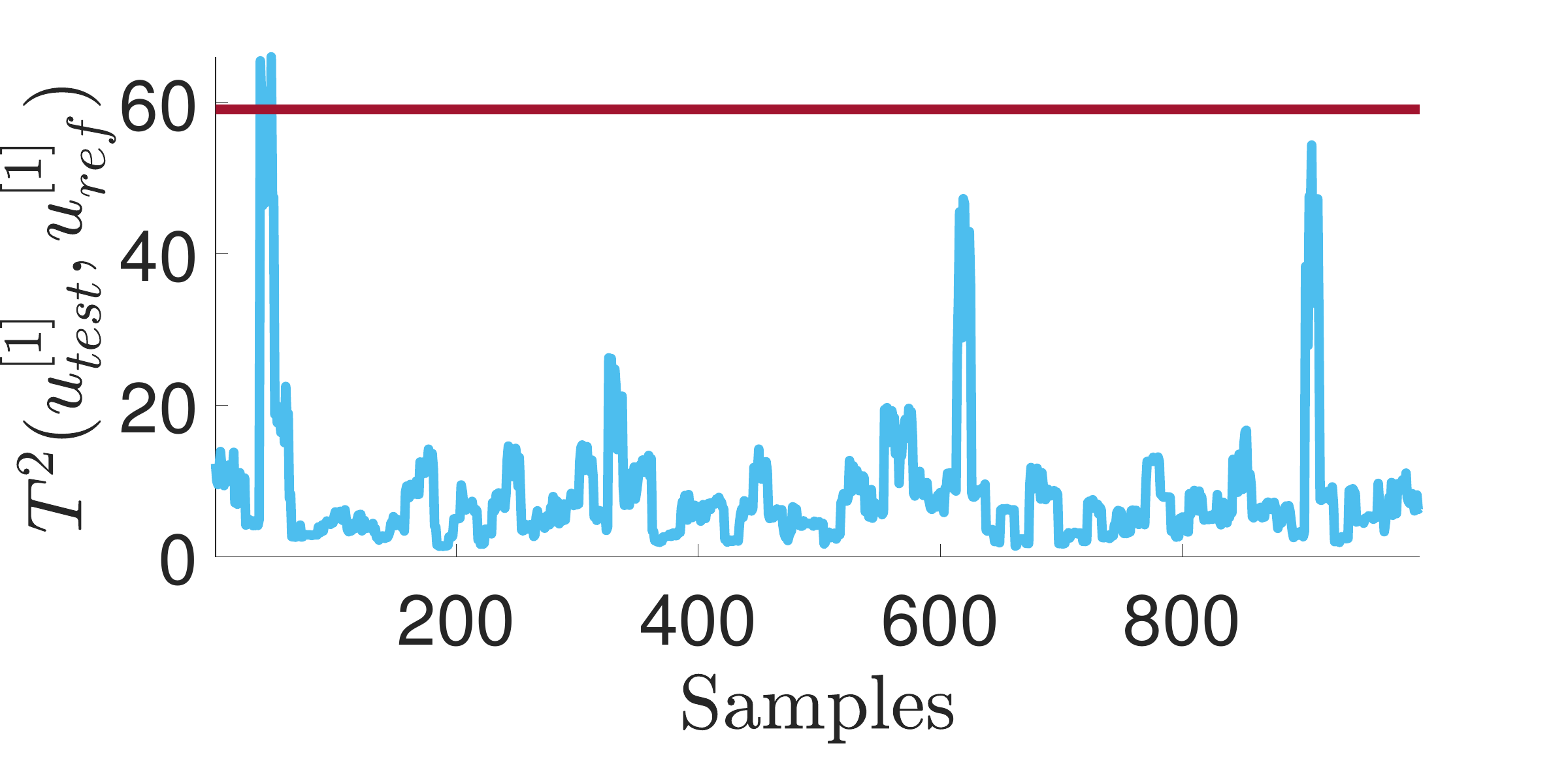} } 
	\caption{ (a) $T_0^s$ used to train $\mathcal{M}^{\scriptscriptstyle[1]}$. (b) Thermal load demands used to train $\mathcal{M}^{\scriptscriptstyle[1]}$, including $P_1^{c}$ (blue), $P_2^{c}$ (purple), $P_3^{c}$ (orange), $P_4^{c}$ (green), $P_5^{c}$ (yellow). (c) Benchmark control chart for the modelling errors of $\mathcal{M}^{\scriptscriptstyle[1]}$, depicting $T^2(\bm{e}_{\text{s},\text{test}},\bm{e}_{\text{s},\text{ref}})$ (light-blue) and UCL$_e$ (red). (d) Benchmark control chart for the inputs used to train $\mathcal{M}^{\scriptscriptstyle[1]}$, depicting $T^2(\bm{u}_\text{test}^{\scriptscriptstyle[1]},\bm{u}_\text{ref}^{\scriptscriptstyle[1]})$ (light-blue) and UCL$_u^{\scriptscriptstyle[1]}$ (red). }
	\label{fig:lowconsCC}
\end{figure}

Then, according to Step 2, a one-week input-output dataset is gathered using a sampling time of $\tau = 5$ minutes. Since the supply temperature $T_0^s$ is controllable, it is varied across its full operating range using multilevel pseudorandom binary sequences (MPRBS), as illustrated in Figure \ref{fig:lowconsCC}(a). The disturbances $P_i^c$ for $i=1,\hdots,5$ follow typical profiles in DHSs \cite{la2023optimal}, and are assumed to be measured during a specific season with thermal demand between 100 and 300 kW (Figure \ref{fig:lowconsCC}(b)). The resulting dataset $\mathcal{D}^{\scriptscriptstyle[1]}$, which also includes the corresponding output variables, is employed to identify the initial model $\mathcal{M}^{\scriptscriptstyle[1]}$, thereby updating the models count to $n=1$. Although various model architectures could be used, in this work we adopt the data-based model proposed in \cite{de2024physics}, which employs gated recurrent unit (GRU) networks within the recurrent neural network (RNN) family. 

Once model $\mathcal{M}^{\scriptscriptstyle[1]}$ is identified, Step 3 of the slow learning procedure involves its statistical characterisation. This includes building the control chart of $T^2(\bm{e}_{\text{s},\text{test}},\bm{e}_{\text{s},\text{ref}})$, see Figure \ref{fig:lowconsCC}(c), containing the control limit UCL$_e=187.5$, computed such that \mbox{$\mathbb{P}_{\!\!e}(T^2(\bm{e}_{\text{s},\text{test}},\bm{e}_{\text{s},\text{ref}}) \leq \text{UCL$_e$})=99.73\%$} (see Section \ref{sec:preliminaries}). Similarly, the control chart of $T^2(\bm{u}_\text{test}^{\scriptscriptstyle[1]},\bm{u}_\text{ref}^{\scriptscriptstyle[1]})$, including UCL$_u^{\scriptscriptstyle[1]}=59.1$, is built and shown in Figure \ref{fig:lowconsCC}(d) to statistically characterise the operating conditions under which $\mathcal{M}^{\scriptscriptstyle[1]}$ has been trained. 
\vspace{2pt}

\begin{figure}
	\centering	\captionsetup[subfloat]{labelfont=scriptsize,textfont=scriptsize}
	\subfloat[]{\includegraphics[width=0.4\textwidth]{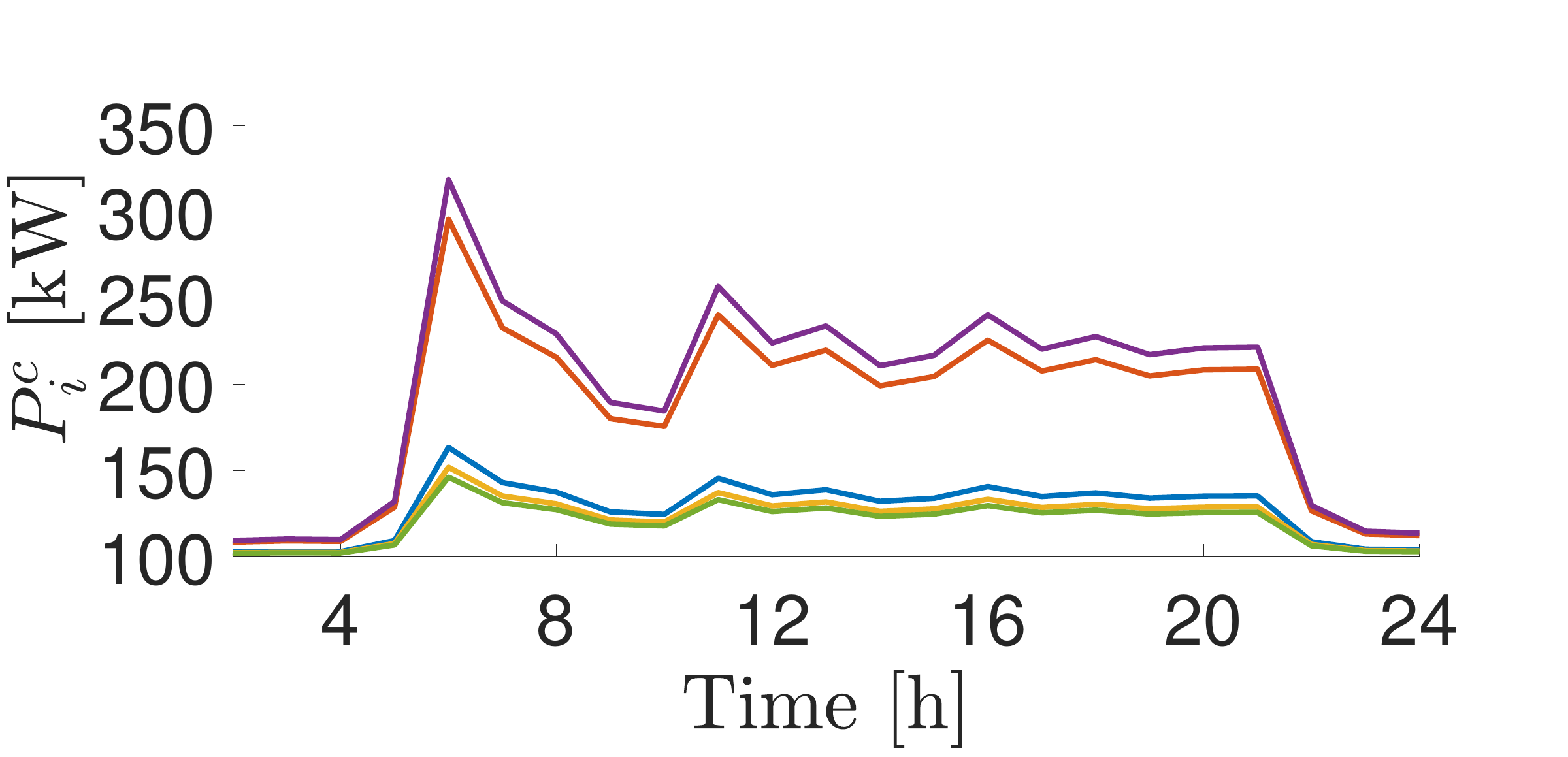} } \\
	\subfloat[]{ \includegraphics[width=0.4\textwidth]{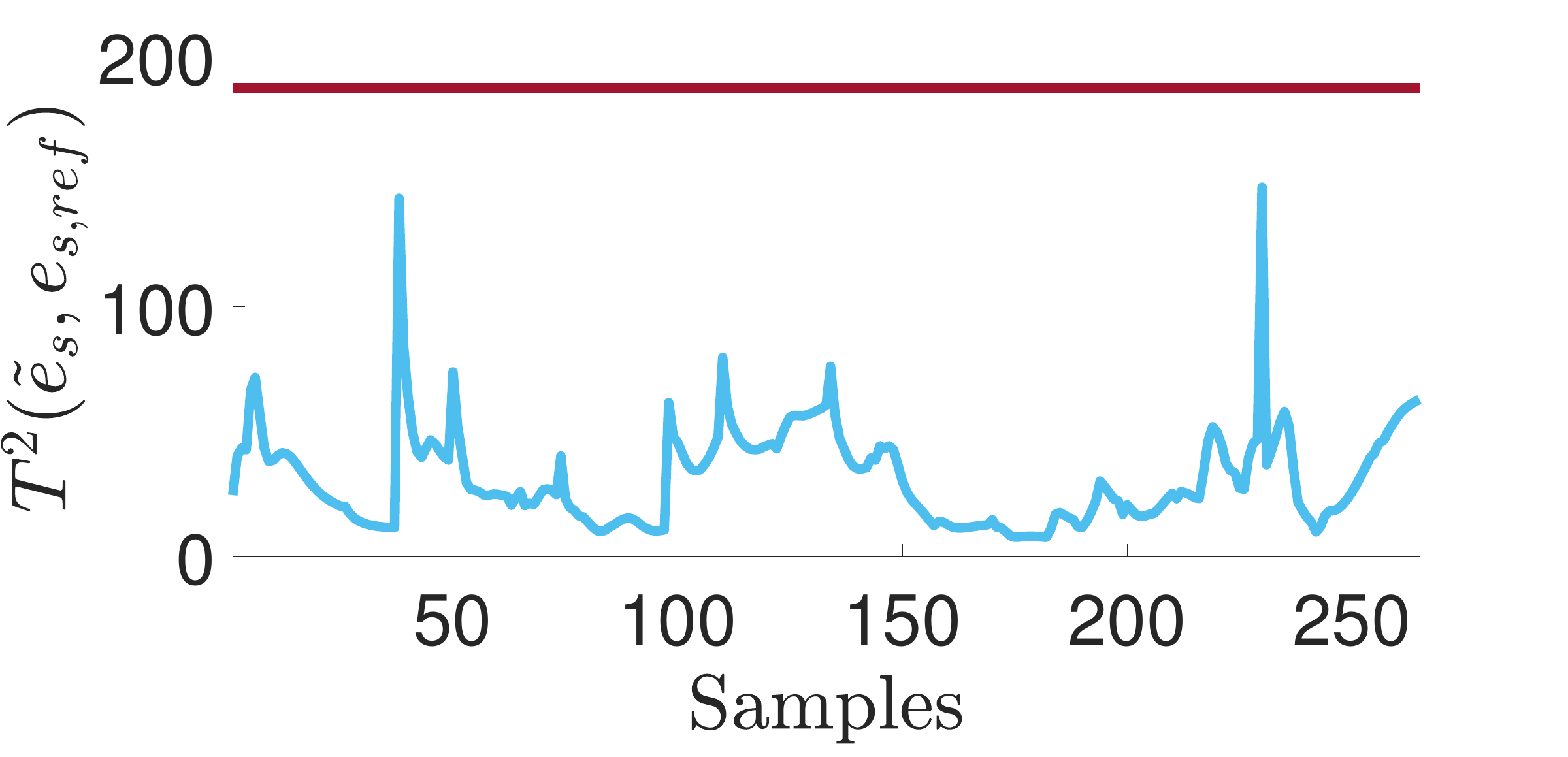} } 
	\subfloat[]{ \includegraphics[width=0.4\textwidth]{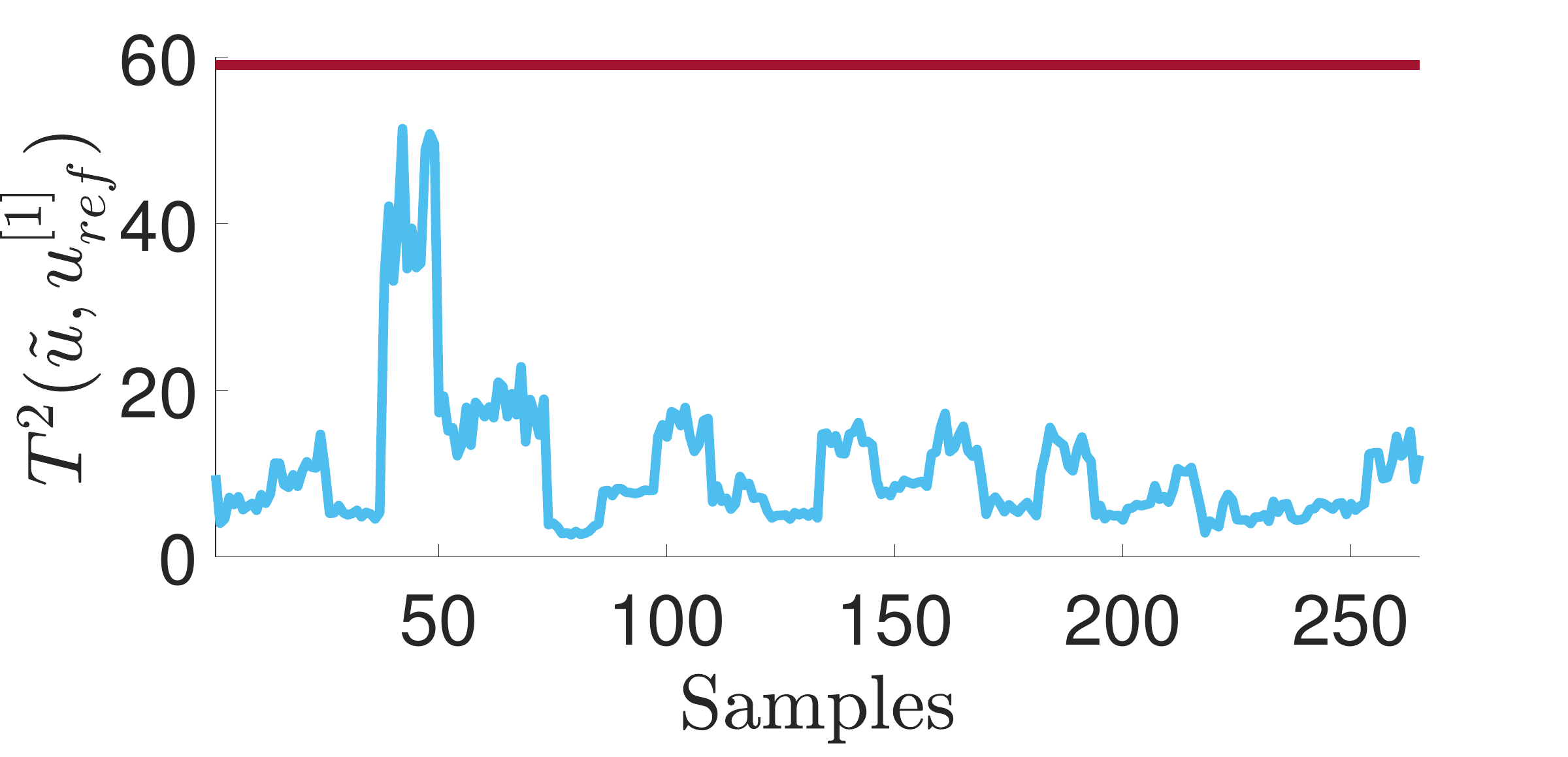} } \\
	\subfloat[]{ \includegraphics[width=0.4\textwidth]{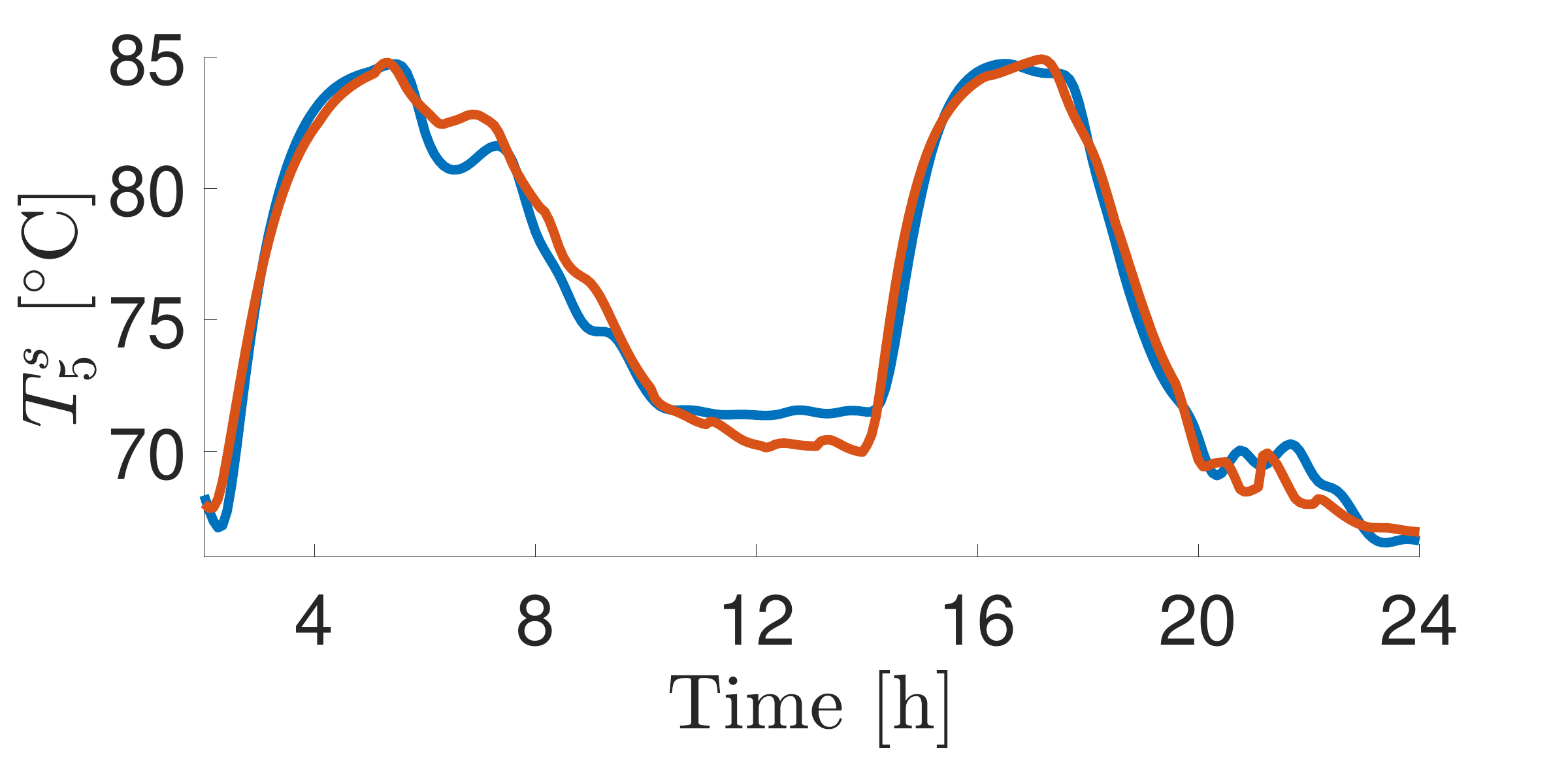} } 
	\subfloat[]{\includegraphics[width=0.4\textwidth]{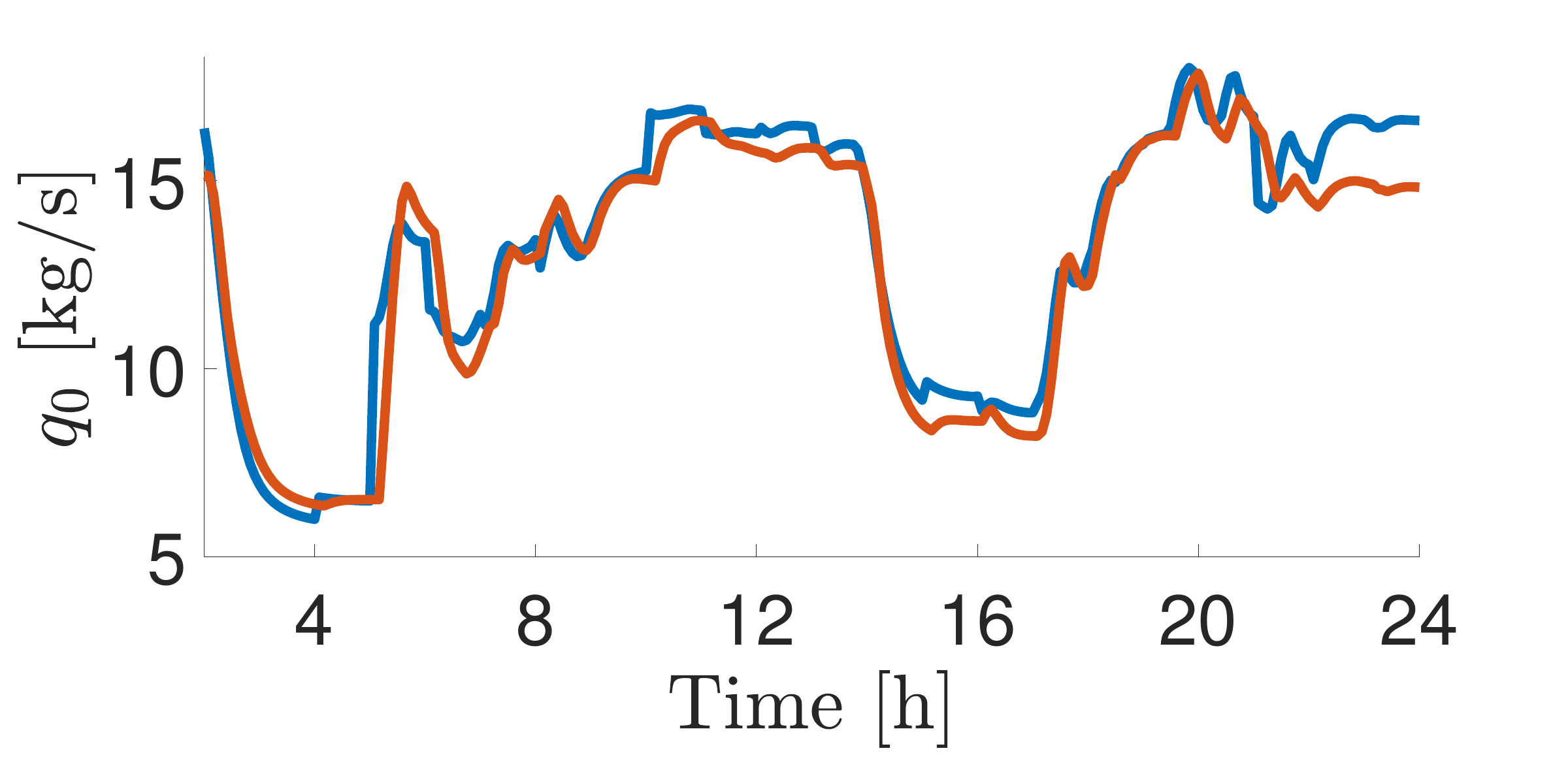} }
	\caption{$\mathcal{M}^{\scriptscriptstyle[1]}$ first test: (a) Thermal load demands test value. (b) Test control chart with respect to the modelling errors of $\mathcal{M}^{\scriptscriptstyle[1]}$, depicting $T^2(\bm{\widetilde{e}}_{\text{s}},\bm{e}_{\text{s}, \text{ref}})$ (light-blue) and UCL$_e$ (red). (c) Test control chart with respect to the inputs used to train  $\mathcal{M}^{\scriptscriptstyle[1]}$, depicting $T^2(\bm{\widetilde{u}}, \bm{u}_\text{ref}^{\scriptscriptstyle[1]})$ (light-blue) and UCL$_u^{\scriptscriptstyle[1]}$ (red). (d) $T_5^s$ predicted by $\mathcal{M}^{\scriptscriptstyle[1]}$ (orange) and measured (blue). (e) $q_0$ predicted by $\mathcal{M}^{\scriptscriptstyle[1]}$ (orange) and measured (blue). }
	\label{fig:lowconslowmodel}
\end{figure}

\textit{\textbf{$\bm{\mathcal{M}^{\scriptscriptstyle[1]}}$ performance monitoring}}: Following Step 4.1 of the slow learning procedure, once model $\mathcal{M}^{\scriptscriptstyle[1]}$ is both identified and statistically characterised, its performance must be continuously monitored to promptly detect anomalies. To this end, two monitoring tests are executed. 

The first daily test, whose samples are collected in the monitoring set $\widetilde{\mathcal{D}}$, evaluates the model performance under input conditions similar to those used for training $\mathcal{M}^{\scriptscriptstyle[1]}$. Specifically, the supply temperature $T_0^s$ varies within a range consistent with the training set, while the disturbance profiles follow the trend shown in Figure \ref{fig:lowconslowmodel}(a), which aligns with the operating domain reported in Figure \ref{fig:lowconsCC}(b). At this point, condition \eqref{eq:cond_e}, which assesses whether the modelling errors are in-control, is checked and, as expected and confirmed by Figure \ref{fig:lowconslowmodel}(b), it is verified. Further verification of the inputs control chart (Figure \ref{fig:lowconslowmodel}(c)) shows that, consistently, condition \eqref{eq:cond_u} is verified, as the current inputs remain statistically close to those used to train $\mathcal{M}^{\scriptscriptstyle[1]}$. This is further validated through visual inspection of key plant variables: Figure \ref{fig:lowconslowmodel}(d) compares the predicted and measured values of $T_5^s$, which must be monitored as it must respect physical constraints being the supply temperature at the most distant load from the heating station \cite{la2023optimal}. Figures \ref{fig:lowconslowmodel}(e) reports the predicted and measured values of $q_0$, whose accurate prediction is crucial to correctly estimate the DHS power consumption. The model achieves a FIT index of 71.2\%, computed as defined in \cite{de2024physics}, indicating satisfactory performance. Therefore, no updates are required, and the algorithm resumes at Step 4. However, despite acceptable results, mismatches between predicted and measured values can be observed. These will be addressed by the fast learning component in Section \ref{subsec:fastlearningresults}.

\begin{figure}
	\centering	\captionsetup[subfloat]{labelfont=scriptsize,textfont=scriptsize}
	\subfloat[]{\includegraphics[width=0.4\textwidth]{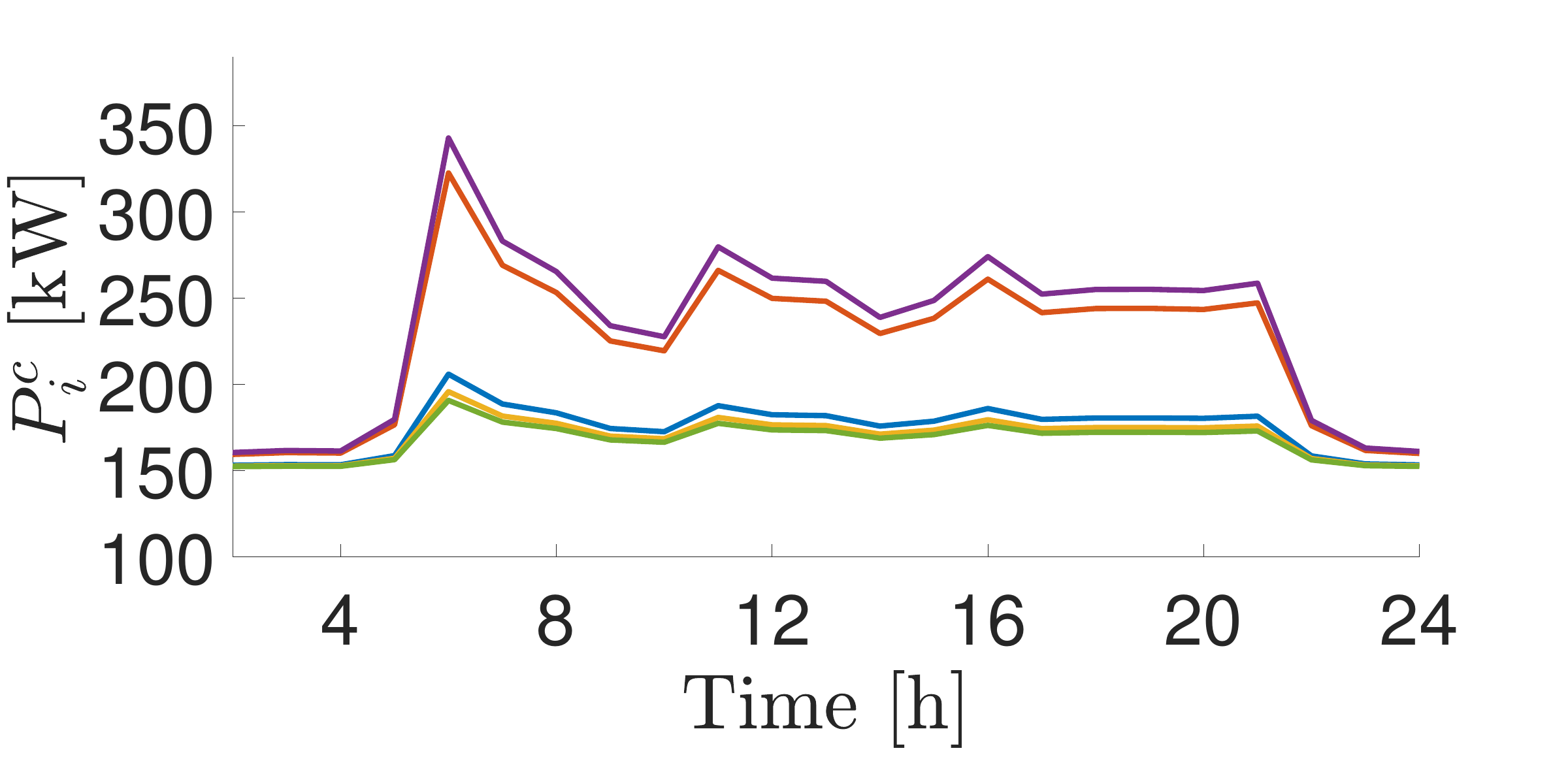} } \\
	\subfloat[]{ \includegraphics[width=0.4\textwidth]{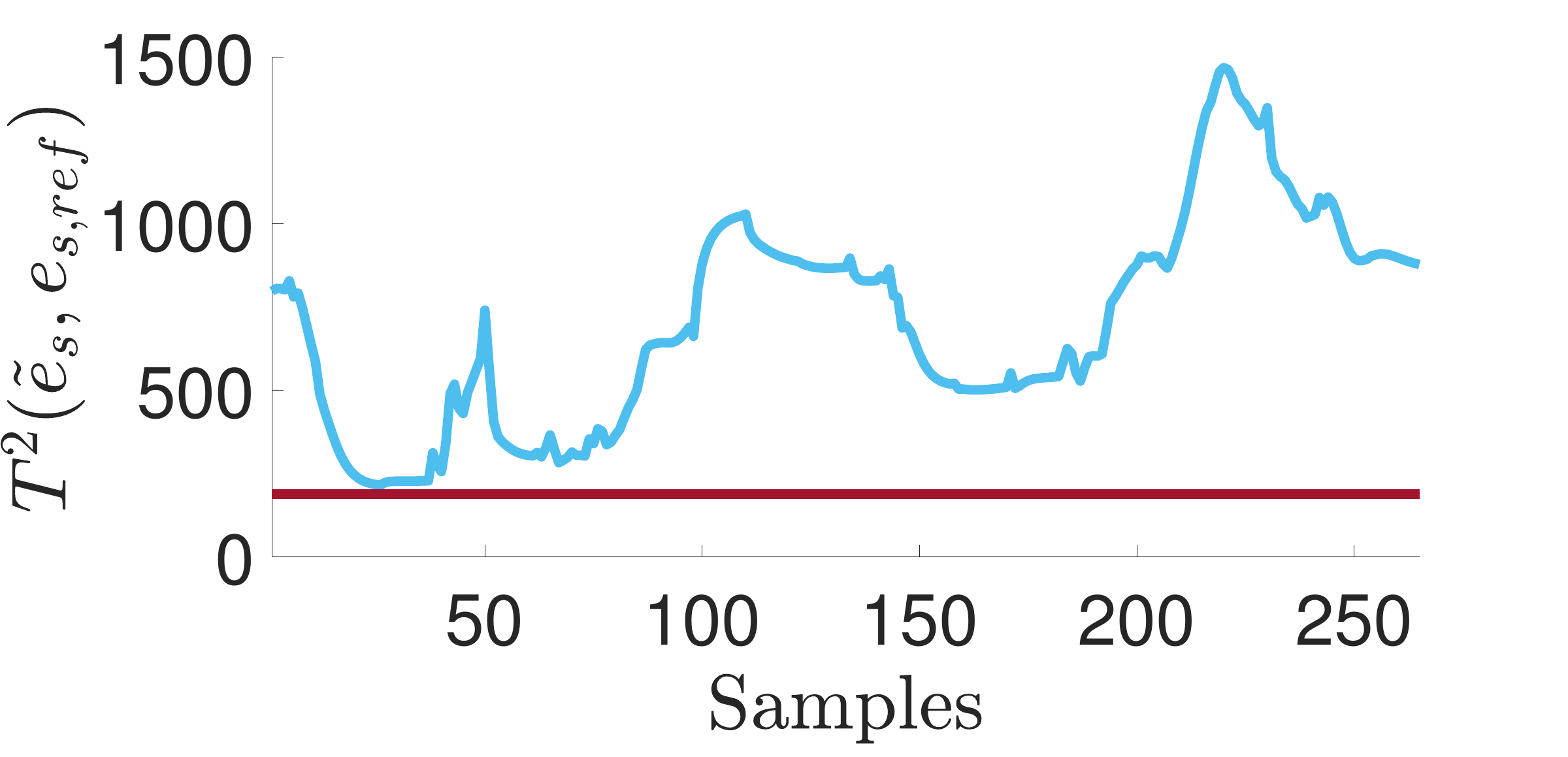} } 
	\subfloat[]{\includegraphics[width=0.4\textwidth]{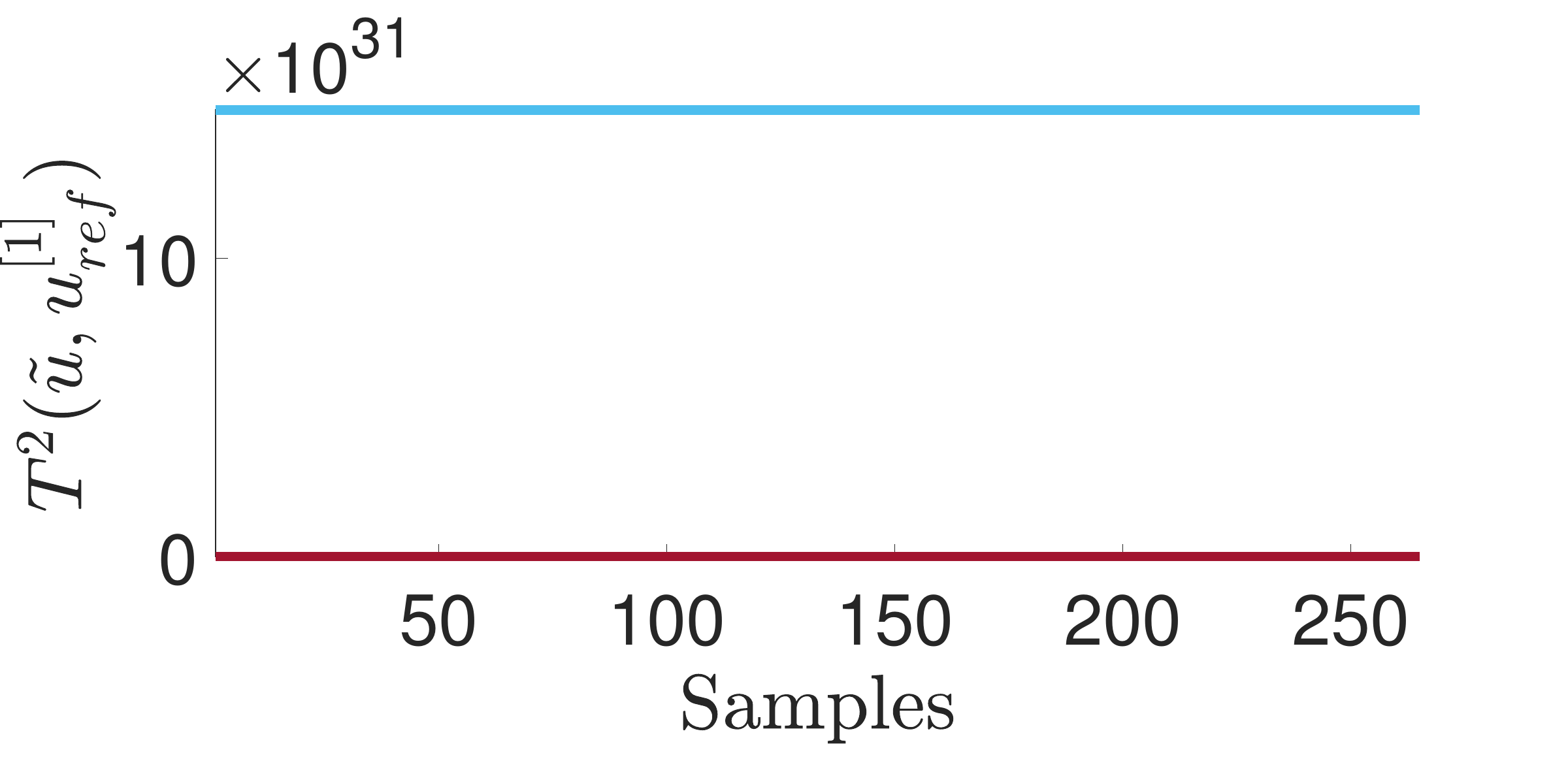} } \\
	\subfloat[]{ \includegraphics[width=0.4\textwidth]{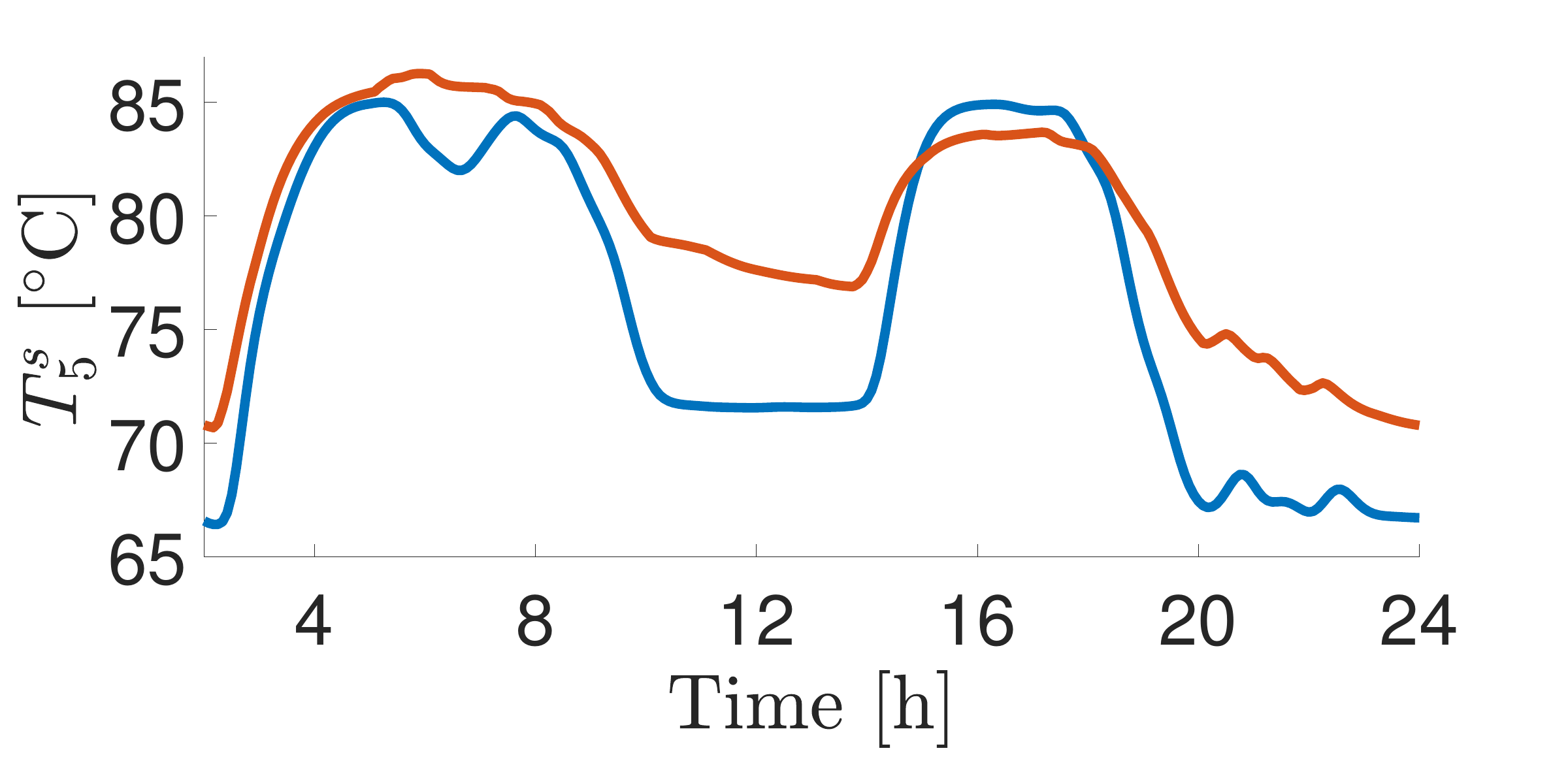} } 
	\subfloat[]{ \includegraphics[width=0.4\textwidth]{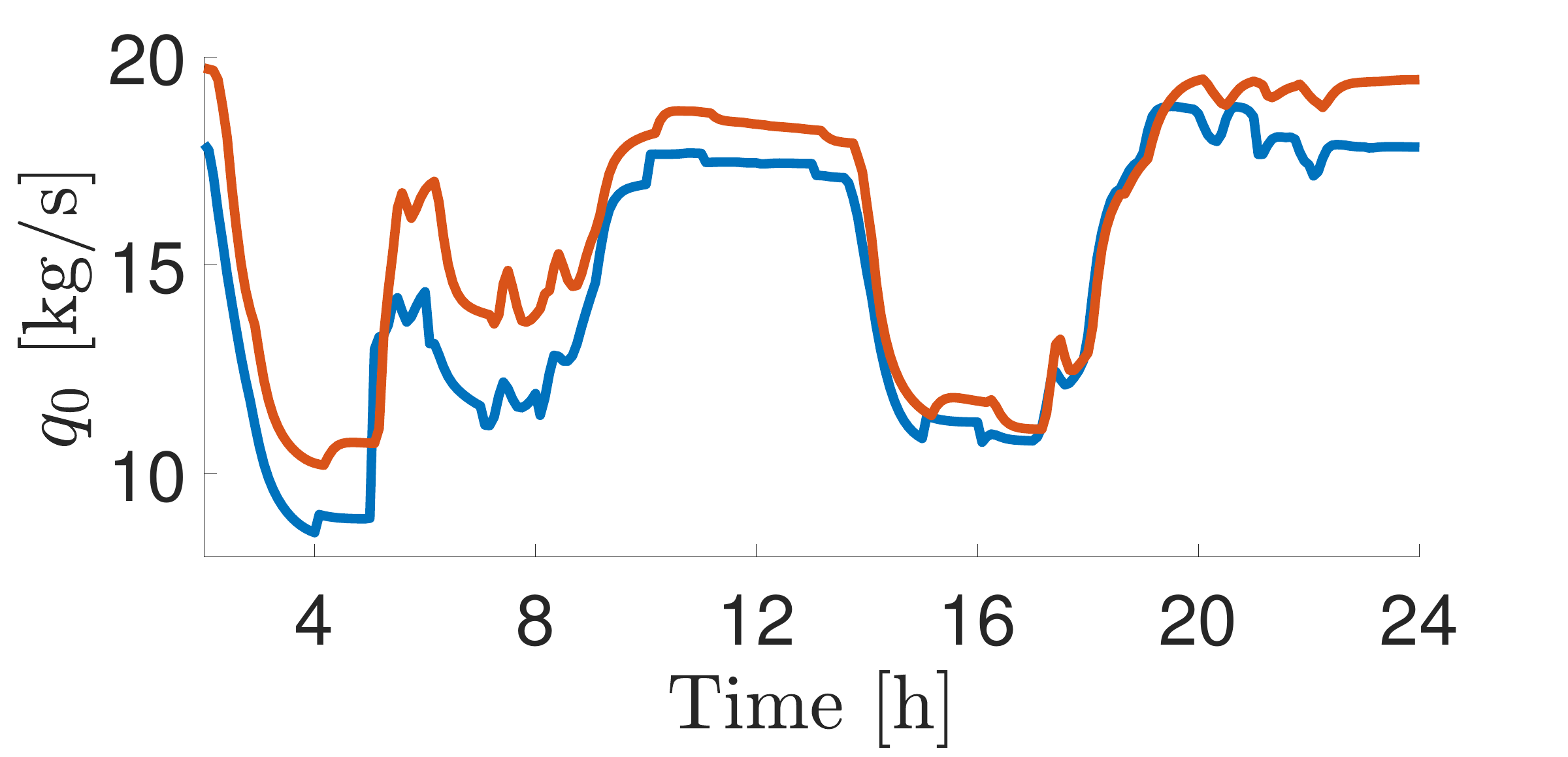} }
	\caption{$\mathcal{M}^{\scriptscriptstyle[1]}$ second test: (a) Thermal load demands test value. (b) Test control chart with respect to the modelling errors of $\mathcal{M}^{\scriptscriptstyle[1]}$, depicting $T^2(\bm{\widetilde{e}}_{\text{s}},\bm{e}_{\text{s}, \text{ref}})$ (light-blue) and UCL$_e$ (red). (c) Test control chart with respect to the inputs used to train $\mathcal{M}^{\scriptscriptstyle[1]}$, depicting $T^2(\bm{\widetilde{u}},\bm{u}_\text{ref}^{\scriptscriptstyle[1]})$ (light-blue) and UCL$_u^{\scriptscriptstyle[1]}$ (red). (d) $T_5^s$ predicted by $\mathcal{M}^{\scriptscriptstyle[1]}$ (orange) and measured (blue). (e) $q_0$ predicted by $\mathcal{M}^{\scriptscriptstyle[1]}$ (orange) and measured (blue). }
	\label{fig:highconslowmodel}
\end{figure}

In a second test, a new monitoring dataset $\widetilde{\mathcal{D}}$ is acquired, keeping the supply temperature within the original training range of $\mathcal{M}^{\scriptscriptstyle[1]}$, but with disturbance profiles now representing an operating condition different from the one in $\mathcal{D}^{\scriptscriptstyle[1]}$, i.e., characterised by higher thermal load demand (150-350 kW), as illustrated in Figure \ref{fig:highconslowmodel}(a). These values may mimic a realistic DHS scenario transitioning from summer to winter conditions, with the latter season typically characterised by higher thermal demands. Following Step 4.1, condition \eqref{eq:cond_e} is no longer satisfied, as witnessed by Figure \ref{fig:highconslowmodel}(b). Consequently, Step 4.2 is invoked to check condition \eqref{eq:cond_u}, which is also found to be violated, as shown in Figure \ref{fig:highconslowmodel}(c). This indicates that $\mathcal{M}^{\scriptscriptstyle[1]}$ fails to correctly capture the system dynamics under this new operating conditions scenario, which in fact was not represented in $\mathcal{D}^{\scriptscriptstyle[1]}$ (see Figures \ref{fig:lowconsCC}(b) and \ref{fig:highconslowmodel}(a)). This conclusion is further supported by a lower FIT of 26.3\% and visual discrepancies between the predictions and measurements shown in Figures \ref{fig:highconslowmodel}(d)-(e). Consequently, the algorithm restores Step 2 to introduce a new model accounting for this newly detected operating condition. 
\vspace{2pt}

\begin{figure}
	\centering	\captionsetup[subfloat]{labelfont=scriptsize,textfont=scriptsize}
	\subfloat[]{ \includegraphics[width=0.4\textwidth]{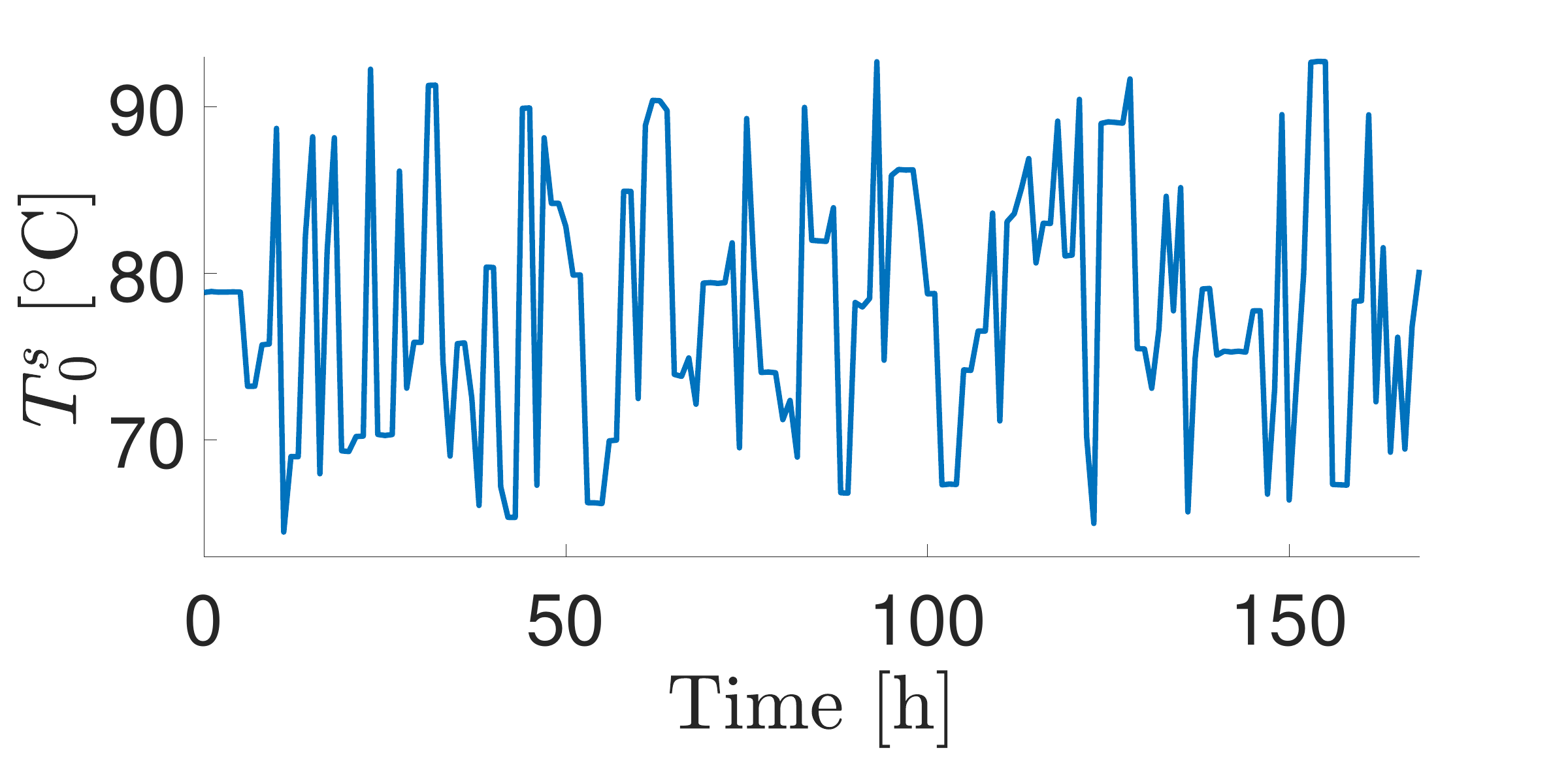} }
	\subfloat[]{\includegraphics[width=0.4\textwidth]{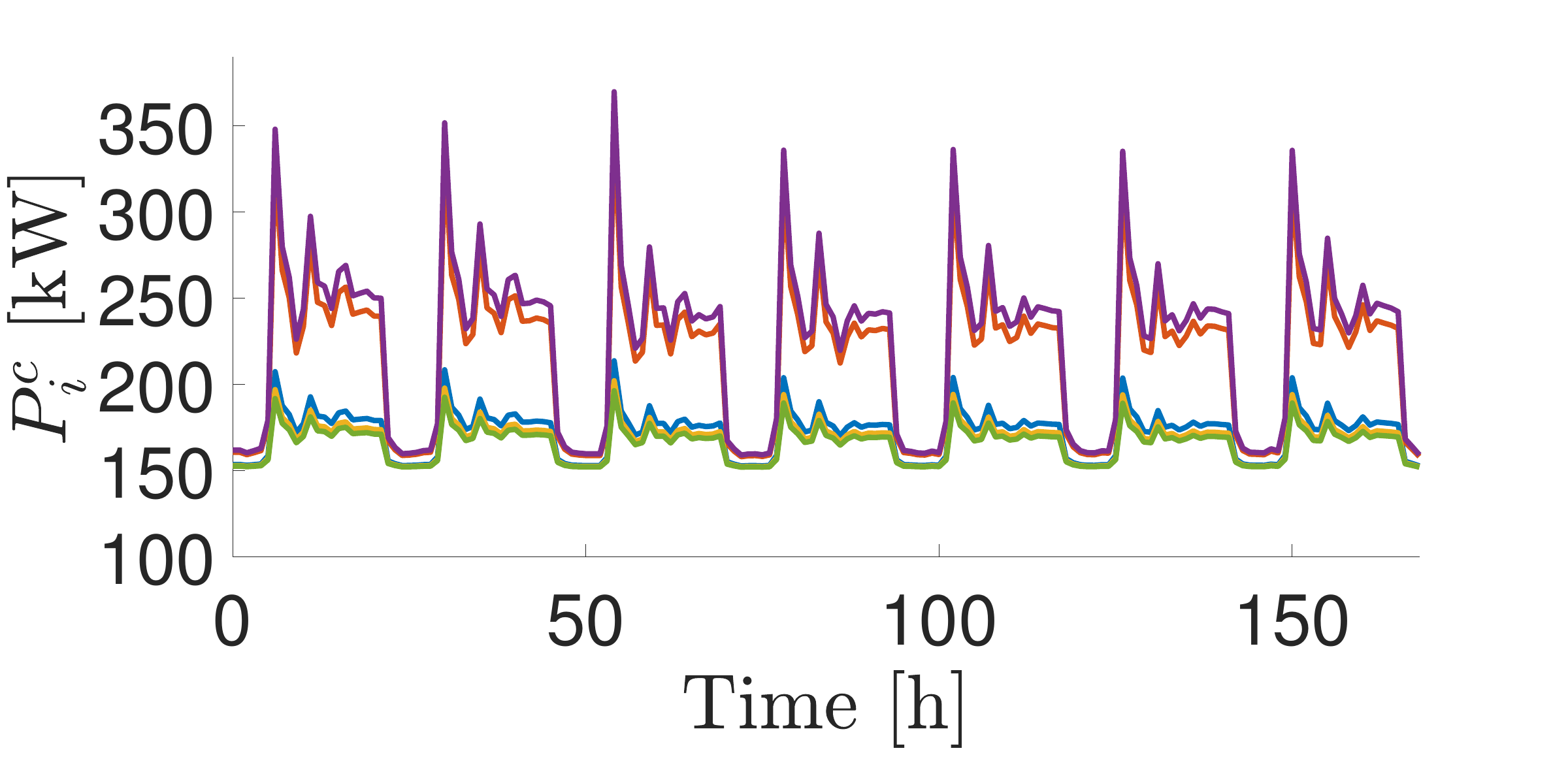} } \\
	\subfloat[]{ \includegraphics[width=0.4\textwidth]{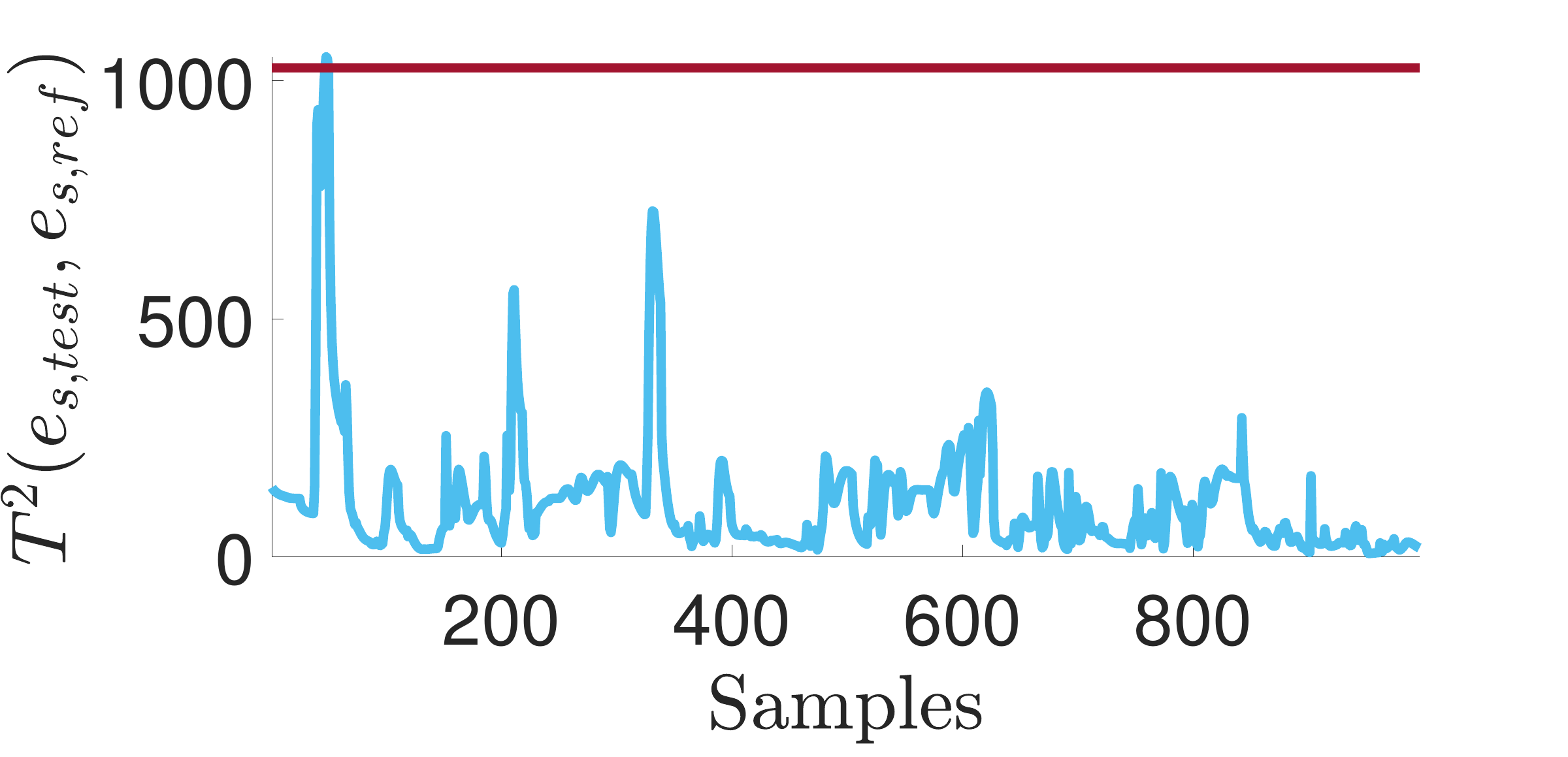} }
	\subfloat[]{\includegraphics[width=0.4\textwidth]{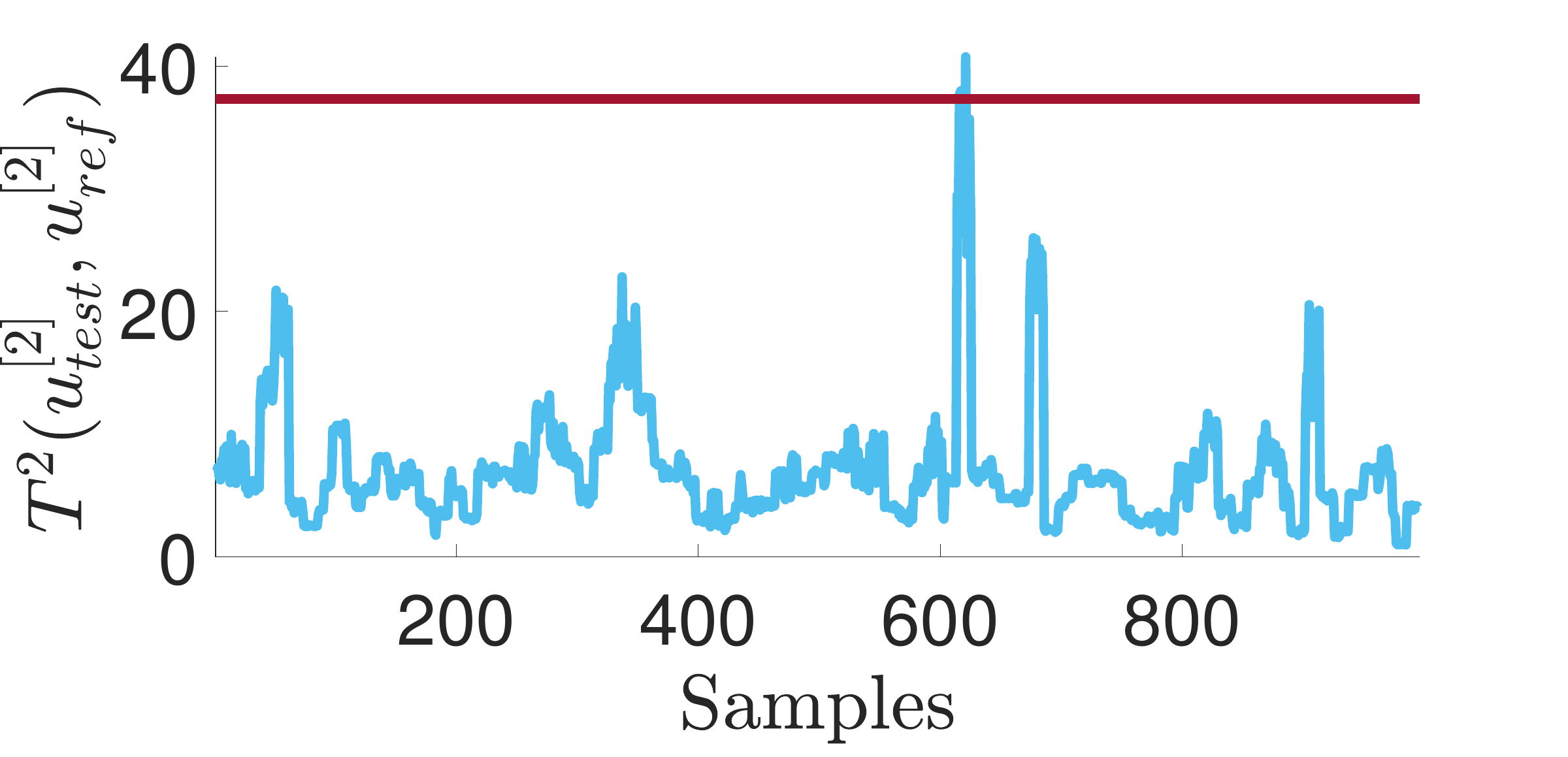} } 
	\caption{ (a) $T_0^s$ used to train $\mathcal{M}^{\scriptscriptstyle[2]}$. (b) Thermal load demands used to train $\mathcal{M}^{\scriptscriptstyle[2]}$. (c) Benchmark control chart for the modelling errors of $\mathcal{M}_{\text{s}}$, depicting $T^2(\bm{e}_{\text{s},\text{test}},\bm{e}_{\text{s},\text{ref}})$ (light-blue) and UCL$_e$ (red). (d) Benchmark control chart for the inputs used to train $\mathcal{M}^{\scriptscriptstyle[2]}$, depicting $T^2(\bm{u}_\text{test}^{\scriptscriptstyle[2]},\bm{u}_\text{ref}^{\scriptscriptstyle[2]})$ (light-blue) and UCL$_u^{\scriptscriptstyle[2]}$ (red). }
	\label{fig:highconsCC}
\end{figure}

\textit{\textbf{Identification of $\bm{\mathcal{M}^{\scriptscriptstyle[2]}}$, integration into the ensemble, and control charts characterisation}}: Following Step 2 of the slow learning procedure, a one-week input-output dataset $\mathcal{D}^{\scriptscriptstyle[2]}$ is collected under the newly encountered operating condition and used to identify $\mathcal{M}^{\scriptscriptstyle[2]}$. The corresponding supply temperature $T_0^s$ remains within the same operating range as before and is displayed in Figure \ref{fig:highconsCC}(a), while the new disturbance profiles are shown in Figure \ref{fig:highconsCC}(b). After identification, $\mathcal{M}^{\scriptscriptstyle[2]}$ is integrated into the ensemble as defined in \eqref{eq:Ms}, yielding $\mathcal{M}_{\text{s}}$. In accordance with Step 3, the control chart for the ensemble modelling errors is constructed by feeding $\mathcal{M}_{\text{s}}$ with the inputs of both $\mathcal{D}^{\scriptscriptstyle[1]}$ and  $\mathcal{D}^{\scriptscriptstyle[2]}$. Thus, $T^2(\bm{e}_{\text{s},\text{test}},\bm{e}_{\text{s},\text{ref}})$ along with the updated control limit UCL$_e=1027.2$ are computed and depicted in Figure \ref{fig:highconsCC}(c), replacing the previous benchmark control chart and limit, i.e., Figure \ref{fig:lowconsCC}(c). Additionally, the control chart for $T^2(\bm{u}_\text{test}^{\scriptscriptstyle[2]},\bm{u}_\text{ref}^{\scriptscriptstyle[2]})$, containing UCL$_u^{\scriptscriptstyle[2]}=37.3$, is built and shown in Figure \ref{fig:highconsCC}(d). This chart is used as benchmark alongside the previously established chart for $T^2(\bm{u}_\text{test}^{\scriptscriptstyle[1]},\bm{u}_\text{ref}^{\scriptscriptstyle[1]})$ (Figure \ref{fig:lowconsCC}(d)) to determine if a new operating condition is encountered. 
\vspace{2pt}

\begin{figure}
	\centering	\captionsetup[subfloat]{labelfont=scriptsize,textfont=scriptsize}
	\subfloat[]{ \includegraphics[width=0.4\textwidth]{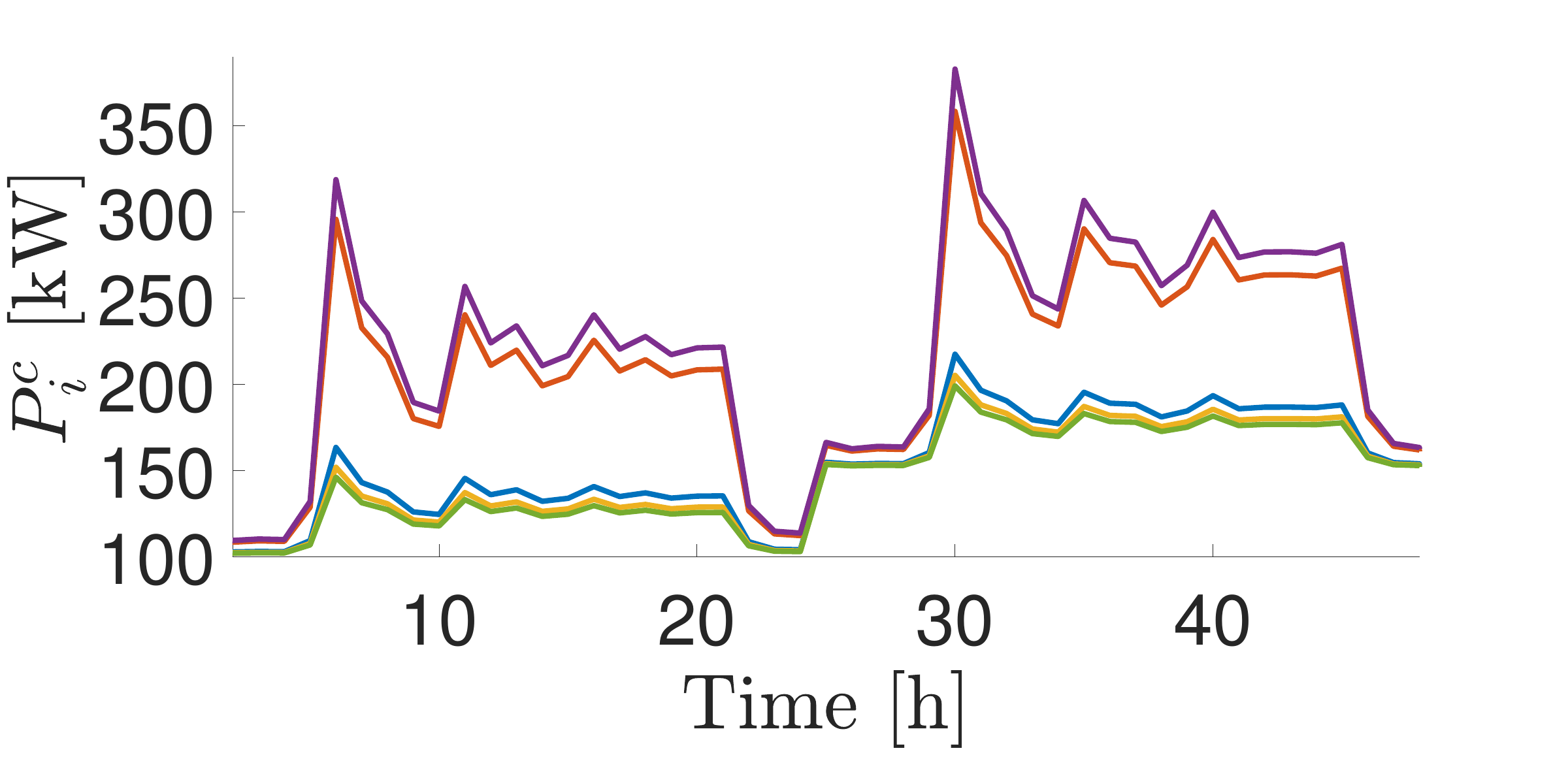} } 
	\subfloat[]{ \includegraphics[width=0.4\textwidth]{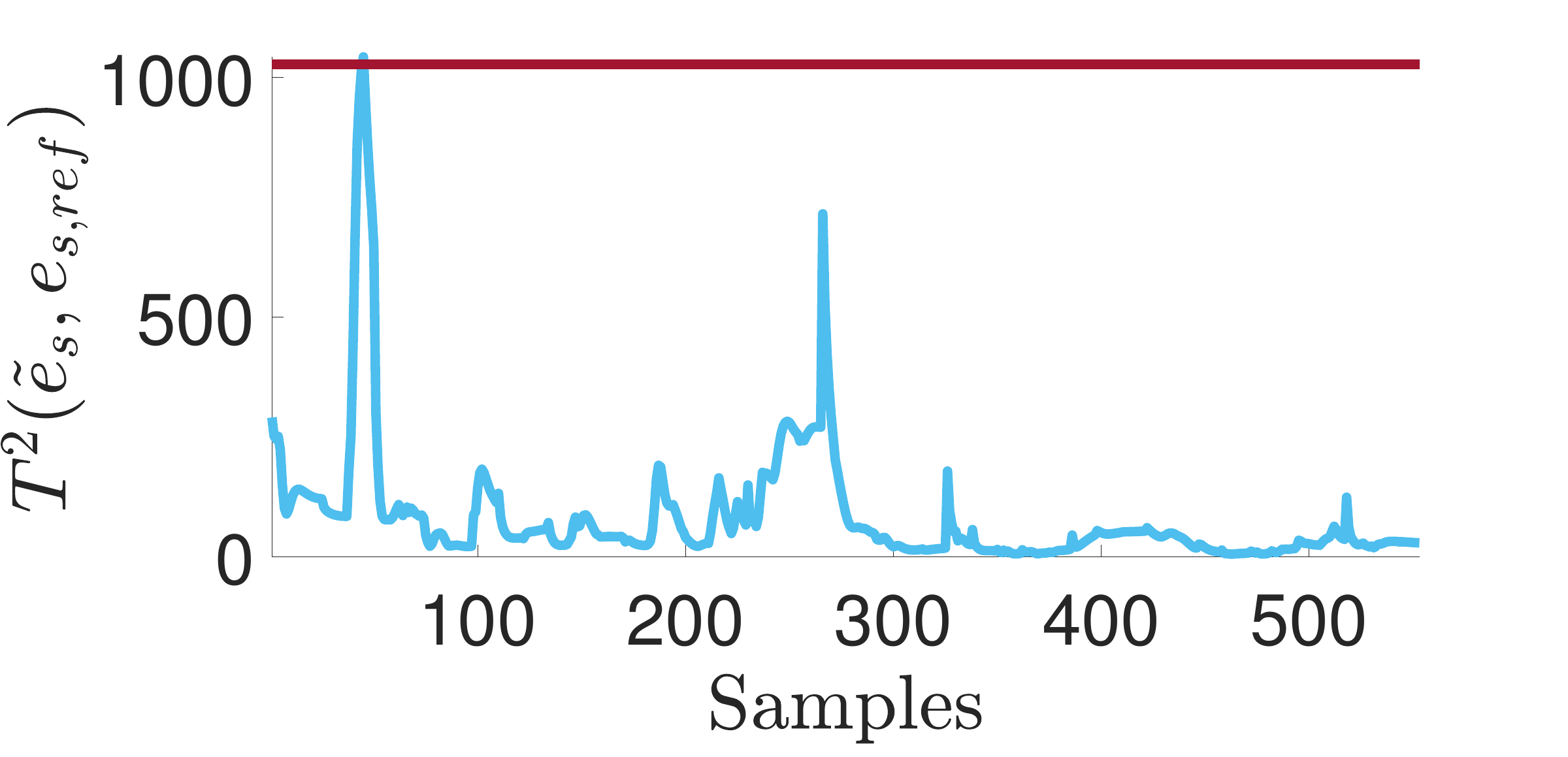} } \\
	\subfloat[]{ \includegraphics[width=0.4\textwidth]{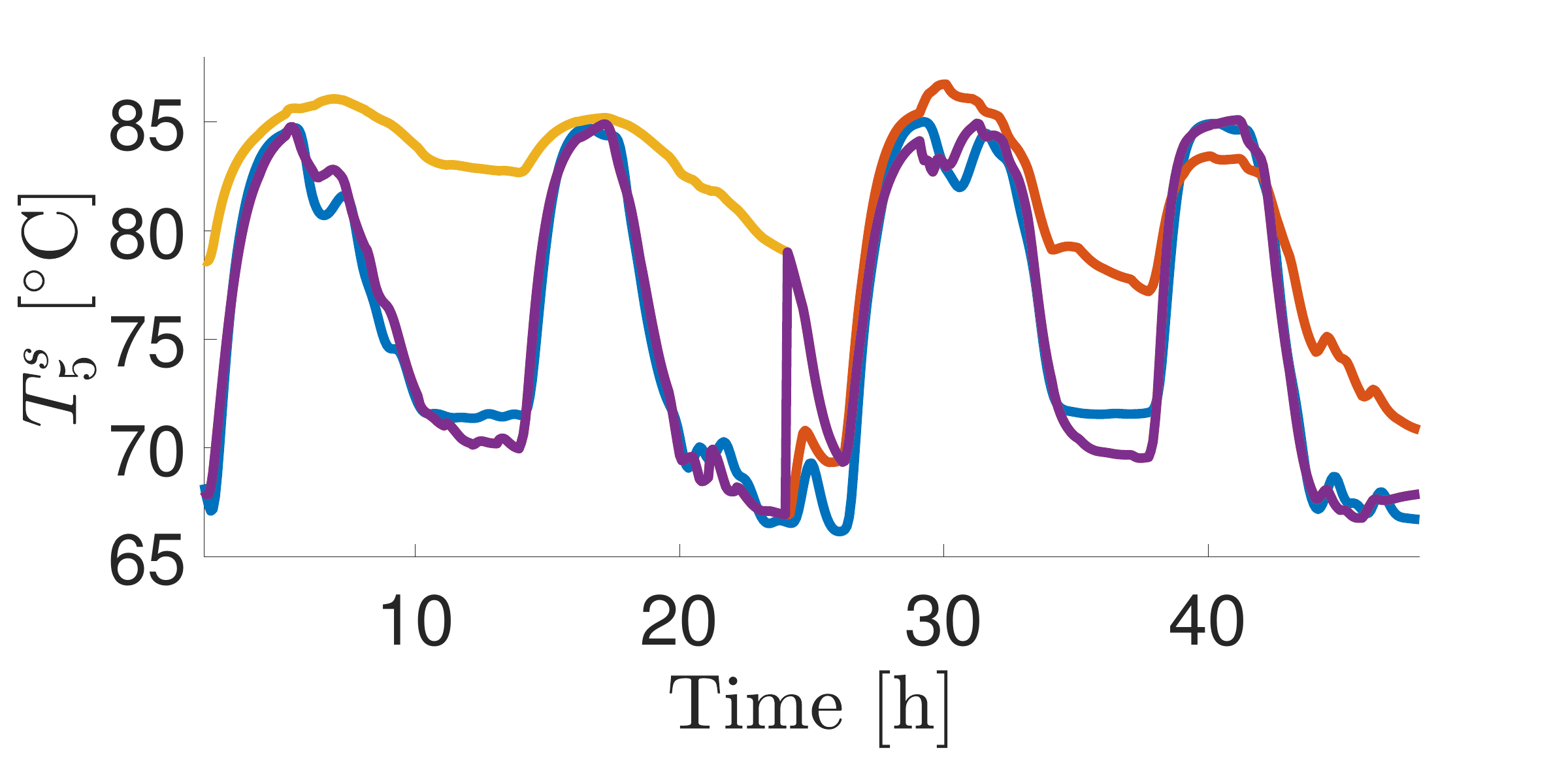} } 
	\subfloat[]{ \includegraphics[width=0.4\textwidth]{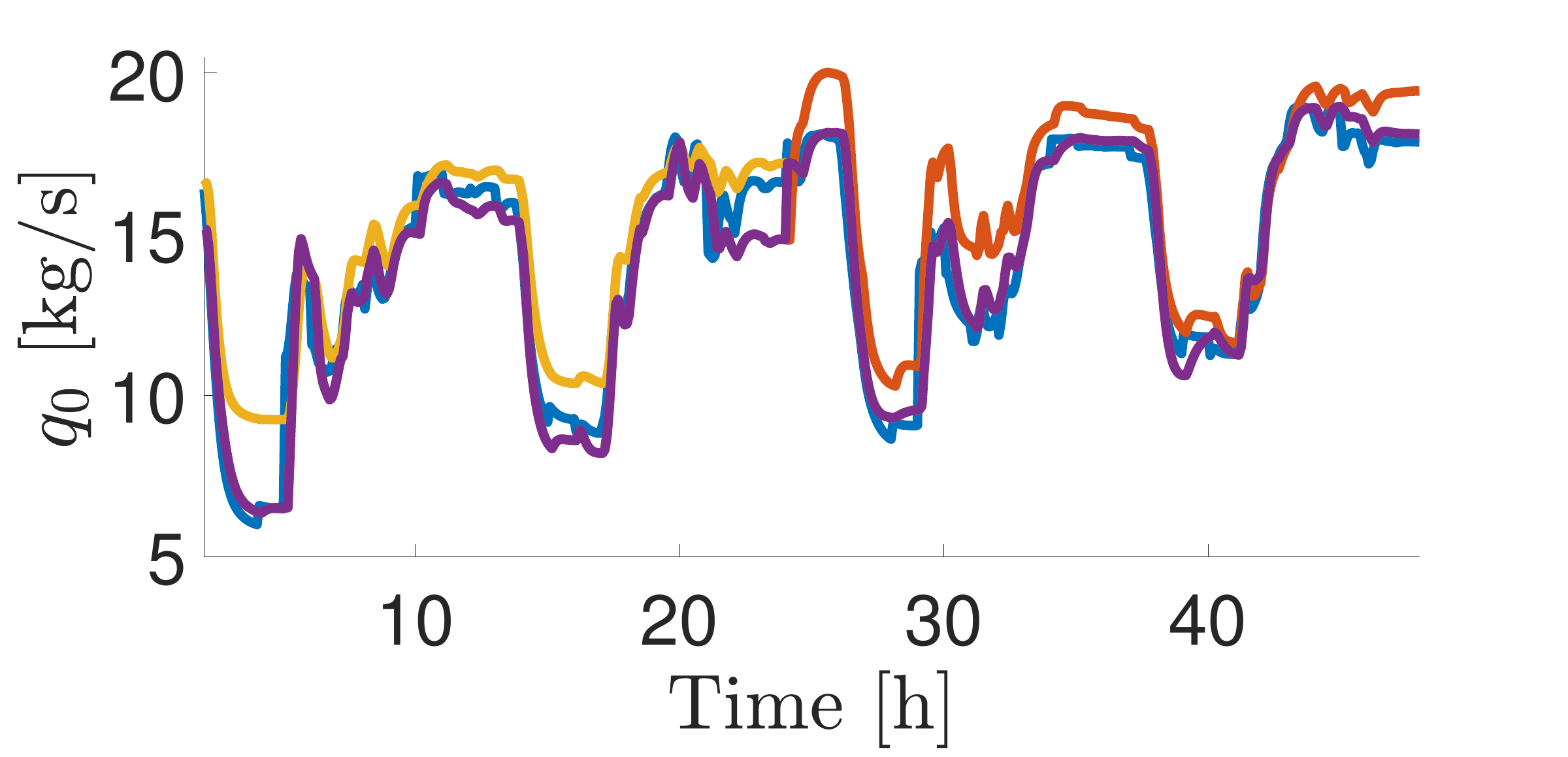} }
	\caption{ Two-day test: (a) Thermal load demands test value. (b) Test control chart with respect to the modelling errors of $\mathcal{M}_{\text{s}}$, depicting $T^2(\bm{\widetilde{e}}_{\text{s}},\bm{e}_{\text{s}, \text{ref}})$ (light-blue) and UCL$_e$. (c) $T_5^s$ predicted by $\mathcal{M}^{\scriptscriptstyle[1]}$ (orange), by $\mathcal{M}^{\scriptscriptstyle[2]}$ (yellow), by $\mathcal{M}_{\text{s}}$ (purple), and measured (blue). (d)  $q_0$ predicted by $\mathcal{M}^{\scriptscriptstyle[1]}$ (orange), by $\mathcal{M}^{\scriptscriptstyle[2]}$ (yellow), by $\mathcal{M}_{\text{s}}$ (purple), and measured (blue). }
	\label{fig:lowhighmodelcomb}
\end{figure}

\textit{\textbf{$\bm{\mathcal{M}_{\text{s}}}$ performance monitoring}}: Following Step 4 of the slow learning procedure, the performance of the ensemble model $\mathcal{M}_{\text{s}}$, combining $\mathcal{M}^{\scriptscriptstyle[1]}$ and $\mathcal{M}^{\scriptscriptstyle[2]}$,  must be monitored to detect potential problems. A two-day test is carried out, collecting the dataset $\widetilde{\mathcal{D}}$: on day one, the disturbance profiles are within the range used to train $\mathcal{M}^{\scriptscriptstyle[1]}$, while on day two, they are within the range used to train $\mathcal{M}^{\scriptscriptstyle[2]}$, as depicted in Figure \ref{fig:lowhighmodelcomb}(a). This configuration enables to test the ensemble model $\mathcal{M}_{\text{s}}$ under two different operating conditions. The control variable remains within its usual operational constraints. Thus, condition \eqref{eq:cond_e} is checked and verified, as depicted in Figure \ref{fig:lowhighmodelcomb}(b), confirming that the rule in \eqref{eq:lambdai} effectively combines the two models. Further validation through visual inspection of key system variables, i.e., $T_5^s$ in Figure \ref{fig:lowhighmodelcomb}(c) and $q_0$ in Figure \ref{fig:lowhighmodelcomb}(d), shows good alignment between predictions and measurements. In particular, from these last plots we can notice that $\mathcal{M}_{\text{s}}$ (purple line) outperforms the individual models (yellow and orange lines) in predicting the real system transients (blue line) across multiple operating conditions. This is further witnessed by the quantitative assessment reported in Table \ref{table:comparisonFS}, which compares the FIT index of $\mathcal{M}^{\scriptscriptstyle[1]}$ alone, $\mathcal{M}^{\scriptscriptstyle[2]}$ alone, and their ensemble $\mathcal{M}_{\text{s}}$ obtained using \eqref{eq:lambdai}. \\
However, as evident from Figures \ref{fig:lowhighmodelcomb}(c)-(d) and the fitting performance of $\mathcal{M}_{\text{s}}$, a persistent plant-model mismatch remains. This will be addressed in real time by the overall fast and slow learning model $\mathcal{M}$, as discussed in the next section.

\subsection{Fast learning results}
\label{subsec:fastlearningresults}
\begin{figure}
	\centering	\captionsetup[subfloat]{labelfont=scriptsize,textfont=scriptsize}
	\subfloat[]{ \includegraphics[width=0.4\textwidth]{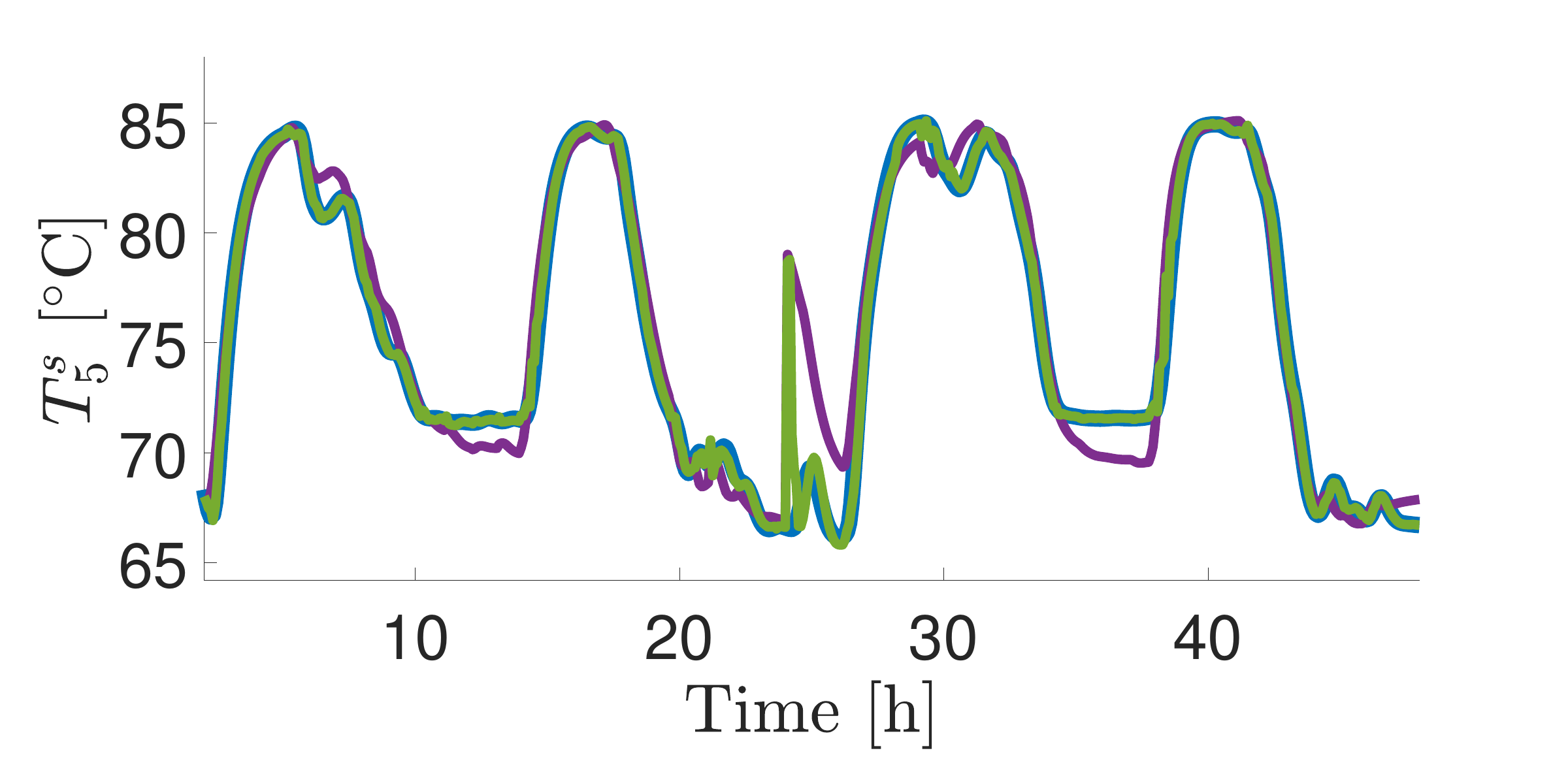} } 
	\subfloat[]{ \includegraphics[width=0.4\textwidth]{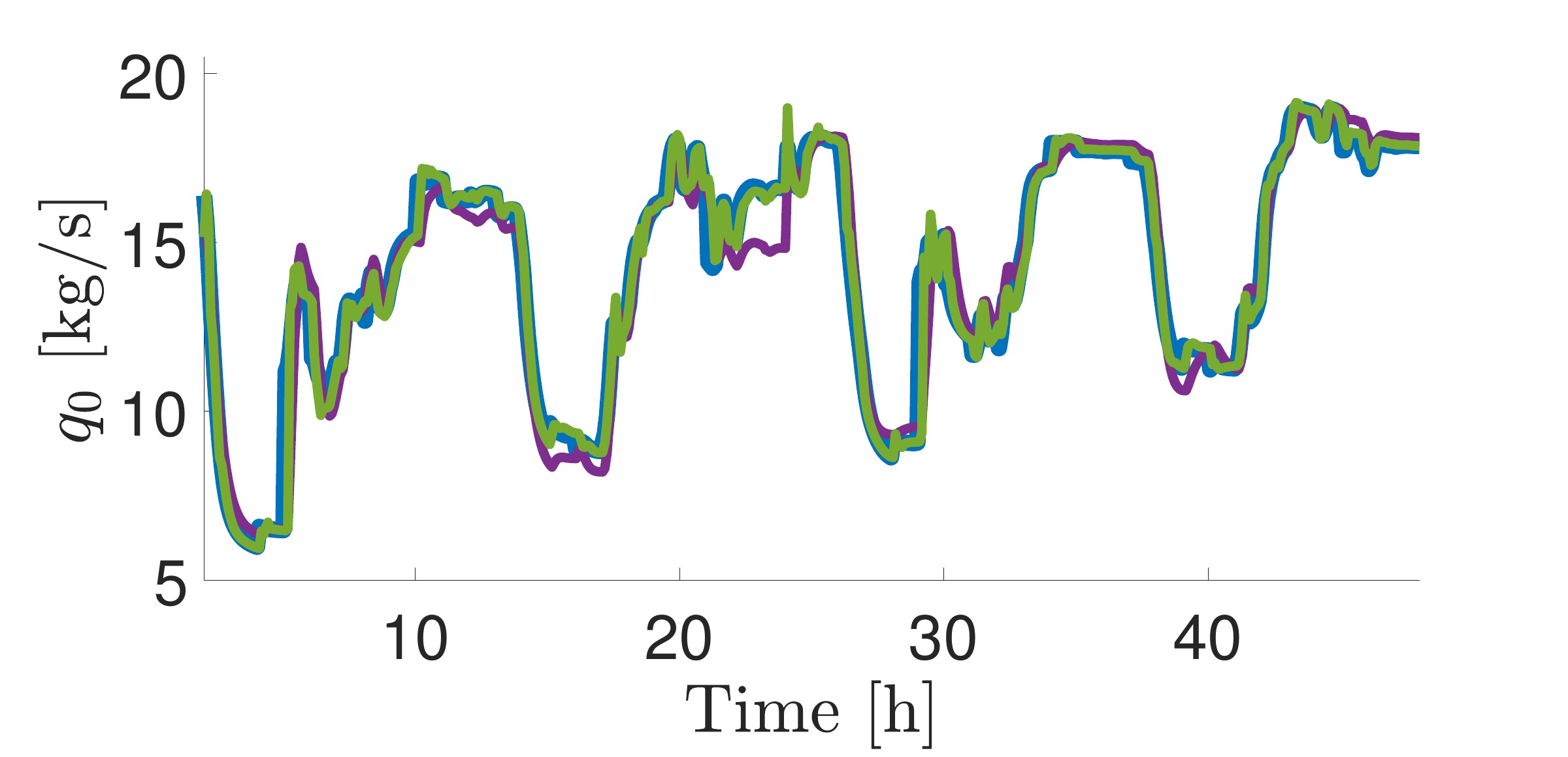} }
	\caption{ Two-day test: (a) $T_5^s$ identified by $\mathcal{M}_{\text{s}}$ (purple), by $\mathcal{M}$ (green), and measured (blue). (b) $q_0$ identified by $\mathcal{M}_{\text{s}}$ (purple), by $\mathcal{M}$ (green), and measured (blue).}
	\label{fig:highmodelcombunc}
\end{figure}

To address the plant-model mismatch of the slow learning component, the architecture presented in Section \ref{sec:procedure} integrates a \textit{fast learned} model $\mathcal{M}_{\text{f}}$ to compensate online for the error of the \textit{slowly learned} model $\mathcal{M}_{\text{s}}$, as detailed in Section \ref{sec:fast}. Specifically, Algorithm \ref{algo:2} is executed every $\tau = 5$ minutes, matching the sampling time of the ensemble model. The selected regression horizons are $n_{r,e} = n_{r,y} = 4$, whereas the minimum and maximum number of samples used for training the GP model are set to $k_{\text{min}}=24$ and $k_{\text{max}}=48$, respectively. This online correction yields the final model output $y$, as defined in \eqref{eq:ytot}, thus enabling the overall model $\mathcal{M}$ to capture the system dynamics more accurately than $\mathcal{M}_{\text{s}}$ alone. The performance improvement achieved through this online correction is illustrated in Figure \ref{fig:highmodelcombunc}, which replicates the test in Figure \ref{fig:lowhighmodelcomb} and compares measured plant variables (blue lines) against the predictions of both $\mathcal{M}_{\text{s}}$ (purple lines) and the corrected model $\mathcal{M}$ (green lines). As visible, the measurements and the predictions of $\mathcal{M}$ are almost overlapping. The enhanced predictive accuracy of $\mathcal{M}$ over $\mathcal{M}_{\text{s}}$ is further confirmed by the FIT index reported in Table \ref{table:comparisonFS}. 

\subsection{Comparison and discussion}
To conclude the analysis, we also compare the individual components with standard modelling techniques. 

First, the slow learning model $\mathcal{M}_{\text{s}}$ is compared with \textit{(i)} a single GRU model $\mathcal{M}^{\scriptscriptstyle[1\text{-}2]}$ trained purely offline on the dataset $\mathcal{D}^{\scriptscriptstyle[1\text{-}2]}=\mathcal{D}^{\scriptscriptstyle[1]}\cup\mathcal{D}^{\scriptscriptstyle[2]}$ covering the full thermal demand range (100-350 kW), and \textit{(ii)} an ensemble $\mathcal{M}_{\scriptscriptstyle\text{AVG}}$ obtained through the arithmetic averaging of $\mathcal{M}^{\scriptscriptstyle[1]}$ and $\mathcal{M}^{\scriptscriptstyle[2]}$. The results, obtained under the same test conditions as in Figure \ref{fig:lowhighmodelcomb} and presented in Table \ref{table:comparisonFS}, show that our $T^2$-based ensemble model $\mathcal{M}_{\text{s}}$ achieves substantially higher fitting performance than both $\mathcal{M}^{\scriptscriptstyle[1\text{-}2]}$ (+65.1\%) and $\mathcal{M}_{\scriptscriptstyle\text{AVG}}$ (+28.5\%).

Second, we compare the slow and fast learning model $\mathcal{M}$ with \textit{(i)} a stand-alone GP model $\mathcal{M}_{\scriptscriptstyle\text{GP}}^{\text{sw}}$ trained exclusively online using the same sliding-window data selection strategy as $\mathcal{M}$ and \textit{(ii)} a stand-alone GP model $\mathcal{M}_{\scriptscriptstyle\text{GP}}^{\text{all}}$ trained purely online without data selection, and which therefore retains all samples. Both models employ the system input $u$ to directly predict the system output $y_p$, following common practice in the literature \cite{maiworm2021online}. The results, obtained under the same test conditions as in Figure \ref{fig:lowhighmodelcomb} and presented in Table \ref{table:comparisonFS}, show that our slow and fast learning model $\mathcal{M}$ outperforms both $\mathcal{M}_{\scriptscriptstyle\text{GP}}^{\text{sw}}$ (+37.0\%) and $\mathcal{M}_{\scriptscriptstyle\text{GP}}^{\text{all}}$ (+53.5\%). 

Finally, it is natural to ask whether the high predictive accuracy achieved by the proposed slow and fast learning model $\mathcal{M}$ comes at the cost of increased computational complexity. To address this aspect, we analyse the average simulation time per iteration $t_{\text{sim}}$, i.e., the average time in seconds required to evaluate each model at a single iteration (5-minute time step) over the two-day simulation scenario. As shown in Table \ref{table:comparisonFS}, purely offline-trained models, namely $\mathcal{M}^{\scriptscriptstyle[1]}$, $\mathcal{M}^{\scriptscriptstyle[2]}$, $\mathcal{M}^{\scriptscriptstyle[1\text{-}2]}$, and $\mathcal{M}_{\scriptscriptstyle\text{AVG}}$, exhibit very low computational times, since only evaluation is required. The model $\mathcal{M}_{\text{s}}$ also belongs to the category of purely offline-trained models, but it entails a slightly higher computational overhead due to the combination rule \eqref{eq:lambdai} which involves the computation of the $T^2$ statistic for each model in the ensemble. Note that if the number of models becomes large, clustering techniques can be employed to limit it. In contrast, purely online-trained models, namely $\mathcal{M}_{\scriptscriptstyle\text{GP}}^{\text{sw}}$ and $\mathcal{M}_{\scriptscriptstyle\text{GP}}^{\text{all}}$, require longer simulation times on average than purely offline-trained models, clearly excluding offline training time. In particular, $\mathcal{M}_{\scriptscriptstyle\text{GP}}^{\text{all}}$ exhibits the highest computational cost, as it is trained online while retaining all collected samples. Ultimately, the proposed two-fold model $\mathcal{M}$ requires slightly more computational time than $\mathcal{M}_{\text{s}}$ and $\mathcal{M}_{\scriptscriptstyle\text{GP}}^{\text{sw}}$, since it integrates both an ensemble of RNNs and a GP. Nevertheless, this modest increase in computational overhead is justified by the substantially higher fitting accuracy achieved by the proposed approach, confirming it to be the best trade-off between fitting performance and computational complexity.

\begin{table}
	\centering
	\setlength{\tabcolsep}{3pt}
	\begin{tabular}{c|cc}
		Model & FIT [\%] & $t_{\text{sim}}$ [s] \\ [0.1cm] \hline  \\[-6px] 
		
		$\mathcal{M}^{\scriptscriptstyle[1]}$ (single RNN trained on $\mathcal{D}^{\scriptscriptstyle[1]}$) & 48.6 & 0.012 \\[2px] 
		
		$\mathcal{M}^{\scriptscriptstyle[2]}$ (single RNN trained on $\mathcal{D}^{\scriptscriptstyle[2]}$) & 42.0 & 0.010 \\[2px] 
		
		$\mathcal{M}_{\text{s}}$ (ensemble with \eqref{eq:lambdai}) & 69.5 & 0.604 \\[2px]
		
		\rowcolor{blue!6} $\mathcal{M}$ (slow and fast learning as in \eqref{eq:ytot}) & 92.9 & 0.829  \\[2px] \hline
		
		$\mathcal{M}^{\scriptscriptstyle[1\text{-}2]}$ (single RNN trained on $\mathcal{D}^{\scriptscriptstyle[1\text{-}2]}$) & 42.1 & 0.020 \\[2px] 
		
		$\mathcal{M}_{\scriptscriptstyle\text{AVG}}$ (ensemble with arithmetic averaging) & 54.1 & 0.021 \\[2px] 
		
		$\mathcal{M}_{\scriptscriptstyle\text{GP}}^{\text{sw}}$ (single sliding-window GP model) & 67.8 & 0.290 \\[2px]  	
		$\mathcal{M}_{\scriptscriptstyle\text{GP}}^{\text{all}}$ (single all-samples GP model) & 60.5 &\!3.451
	\end{tabular}
	\caption{Fitting performance and computational time of the different models tested in the work for a two-day experiment. }
	\label{table:comparisonFS}
\end{table}

\section{Conclusions}
\label{sec:conclusions}
This article proposes a novel machine learning framework for adapting data-based models over time through a two-fold architecture. The \textit{slow learning} component incrementally learns new process dynamics using an ensemble of models, where \textit{i)} each model output is weighted according to the statistical proximity of its training data to the current operating condition and \textit{ii)} a new model is added to the ensemble only when a control chart-based strategy detects that the existing ensemble has become unreliable due to a shift in operating conditions. The \textit{fast learning} component employs a Gaussian process to continuously compensate in real time for the mismatch of the slow learning model arising from real-world variability. The proposed framework is evaluated on a district heating system characterised by multiple operating conditions and persistent plant-model mismatch, showing that the ensemble combination rule outperforms both individual RNNs and simple averaging strategies and that the monitoring algorithm ensures that new models are added to the ensemble in response to new operating conditions. Furthermore, by correcting the mismatch of the slow learning component online at each iteration, the fast component significantly enhances model accuracy, resulting in a high-performing and reliable overall model that adapts effectively to system changes over time. 

The proposed framework presents some limitations that may motivate future research directions. First, the high predictive accuracy of the proposed modelling framework may affect its computational efficiency. 
This can be solved by clustering models associated with close operating conditions in the slow learning component, and by employing more efficient online learning approaches for the fast learning component, e.g., Bayesian neural networks \cite{jospin2022hands}.  
Moreover, considering that adaptation algorithms improve modelling accuracy in the presence of informative data, future work will focus on integrating the proposed modelling architecture into an active learning-based model predictive controller that steers the system to collect informative data while satisfying safety constraints.

\section*{Acknowledgments}
The authors gratefully acknowledge Giuseppe De Nicolao for the fruitful discussions. The work was carried out within the MICS (Made in Italy - Circular and Sustainable) Extended Partnership and received funding from Next-Generation EU (Italian PNRR - M4 C2, Invest 1.3 - D.D. 1551.11-10-2022, PE00000004). CUP MICS D43C22003120001.

\bibliographystyle{IEEEtran}
\bibliography{Bibliography}

\end{document}